\documentclass[prx,10pt,a4paper,twocolumn,nofootinbib, superscriptaddress]{revtex4-1}
\usepackage{graphicx}
\usepackage{amsmath}
\usepackage{amsfonts}
\usepackage{amssymb}
\usepackage{amstext} 
\usepackage{soul}
\usepackage{dsfont}
\usepackage[utf8]{inputenc}
\usepackage[hidelinks]{hyperref}
\usepackage[caption=false]{subfig}
\usepackage{epstopdf} 
\usepackage{cellspace}
\hypersetup{colorlinks=true, citecolor=blue, urlcolor=blue, linkcolor=blue}
\usepackage{tabularx,ragged2e,booktabs}
\usepackage[dvipsnames]{xcolor}
\usepackage{braket}

\newcommand{\A}{\nonumber}
\newcommand{\RN}[1]{\textup{\uppercase\expandafter{\romannumeral#1}}}

\renewcommand{\rm}{\mathrm}

\date{\today}

\begin{document}

\title{Optimal estimation with quantum optomechanical systems in the nonlinear regime}

\author{Fabienne Schneiter}
\affiliation{Institut f\"{u}r Theoretische Physik, Eberhard-Karls-Universit\"{a}t T\"{u}bingen, D-72076 T\"{u}bingen, Germany}

\author{Sofia Qvarfort}
\email{sofiaqvarfort@gmail.com}
\affiliation{Department of Physics and Astronomy, University College London, Gower Street, WC1E 6BT London, United Kingdom}

\author{Alessio Serafini}
\affiliation{Department of Physics and Astronomy, University College London, Gower Street, WC1E 6BT London, United Kingdom}

\author{Andr\'{e} Xuereb} 
\affiliation{Department of Physics, University of Malta, Msida MSD 2080, Malta}

\author{Daniel Braun}
\affiliation{Institut f\"{u}r Theoretische Physik, Eberhard-Karls-Universit\"{a}t T\"{u}bingen, D-72076 T\"{u}bingen, Germany}

\author{Dennis R\"atzel}
\email{dennis.raetzel@physik.hu-berlin.de}
\affiliation{Institut f\"ur Physik, Humboldt Universit\"at zu Berlin, Newtonstra\ss e 15, 12489 Berlin}

\author{David Edward Bruschi}
\email{david.edward.bruschi@gmail.com}
\affiliation{Faculty of Physics, University of Vienna, Boltzmanngasse 5, 1090 Vienna, Austria}
\affiliation{Institute for Quantum Optics and Quantum Information - IQOQI Vienna, Boltzmanngasse 3, 1090 Vienna, Austria}
\affiliation{Theoretical Physics, Universit\"at des Saarlandes,  66123 Saarbr\"ucken, Germany}

\begin{abstract}
\noindent We study the fundamental bounds on precision measurements of parameters contained in a time-dependent nonlinear optomechanical Hamiltonian, which includes the nonlinear light--matter coupling, a mechanical displacement term, and a single-mode mechanical squeezing term. By using a recently developed method to solve the dynamics of this system, we derive a general expression for the quantum Fisher information and demonstrate its applicability through three concrete examples: estimation of the strength of a nonlinear light--matter coupling, the strength of a time-modulated mechanical displacement, and a  single-mode mechanical squeezing parameter, all of which are modulated at resonance. Our results can be used to compute the sensitivity of a nonlinear optomechanical system to a number of external and internal effects, such as forces acting on the system or modulations of the light--matter coupling. 
\end{abstract}

\maketitle

\section{Introduction}\label{intro}
Quantum metrology is the study of sensing schemes that make use of unique properties of quantum systems, such as coherence and entanglement~\cite{paris09}. Sensing with quantum systems is generally superior compared with classical schemes since these quantum properties fundamentally alter the rate at which information can be acquired~\cite{giovannetti2006quantum}. 

A key task within the study of quantum metrology entails  investigating the sensing capabilities that can be achieved with different quantum systems. 
Quantum sensing now features prominently in the planning and building of larger-scale experimental efforts, such as the inclusion of squeezed light in the Advanced Laser Interferometer Gravitational-Wave Observatory (LIGO)~\cite{Aasi:2013} and space-based tests of microgravity~\cite{van2010bose}. Additional prominent candidates for quantum sensors include atomic and molecular interferometers for accelerometry and rotation measurements~\cite{lenef1997rotation}. Similarly, Bose-Einstein condensates have been proposed as platforms for testing fundamental physics~\cite{bruschi2014testing, howl2019exploring} and precision measurements of external potentials~\cite{ratzel2019testing}. Quantum advantages in sensing are also furthering the emergence of quantum precision technologies ~\cite{wisemanbook}, which include atomic clocks~\cite{giovannetti2011advances} and extremely precise magnetic-field sensors~\cite{liu2019nanoscale, fiderer2018quantum}. 

Optomechanical systems~\cite{marquardt2009optomechanics}, which consist of a mechanical element interacting with light, have emerged as ideal candidates for a number of sensing applications~\cite{arcizet2006high}. Due to the large mass of the mechanical element, many proposals in fundamental physics could potentially be tested with optomechanical experiments, such as collapse theories~\cite{PhysRevLett.112.210404, nimmrichter2014optomechanical, pfister2016universal}. Furthermore, optomechanical systems have been proposed as the main experimental platform for detection of possible low-energy quantum gravity effects~\cite{bose2017spin, marletto2017gravitationally, belenchia2018quantum}. In terms of force sensing, microspheres optically trapped in a lattice have been considered~\cite{ranjit16,rashid17}, as well as mesoscopic interferometry for the purpose of gravitational wave detection~\cite{marshman2018mesoscopic}. 

The addition of a cavity to the optomechanical system introduces an inherently nonlinear cubic  interaction between the electromagnetic field and the mechanical element~\cite{Aspelmeyer:Kippenberg:2014}. For systems operating in the nonlinear regime, the quantum Fisher information (QFI) for  measurements of constant gravitational acceleration  has already been computed~\cite{Qvarfort:Serafini:2018, armata2017quantum}, and optimal estimation schemes for the nonlinear coupling itself have been considered~\cite{bernad2018optimal}. In general, the estimation of anharmonicities present in the system  is a topic of great interest~\cite{rivas10,latmiral16} as well as the enhancement of parameter estimation granted by Kerr nonlinearities~\cite{genoni09,rossi16}. Additional efforts have focused on parametric driving of the cavity frequency, which manifests itself as a single-mode mechanical squeezing term in the Hamiltonian~\cite{farace2012enhancing}.

To date, due to challenges in solving the dynamical evolution for time-dependent nonlinear  optomechanical systems, most approaches to the full nonlinear case have been restricted to the estimation of static effects. As a result, the proposals considered so far are of limited interest for experimentalists, since static effects are generally difficult to isolate from a random noise floor. Furthermore, if feasible, time-dependent signals also allow for the exploitation of resonances, which can be used to increase the signal-to-noise ratio.

\begin{figure}[t!]
\includegraphics[width=\linewidth]{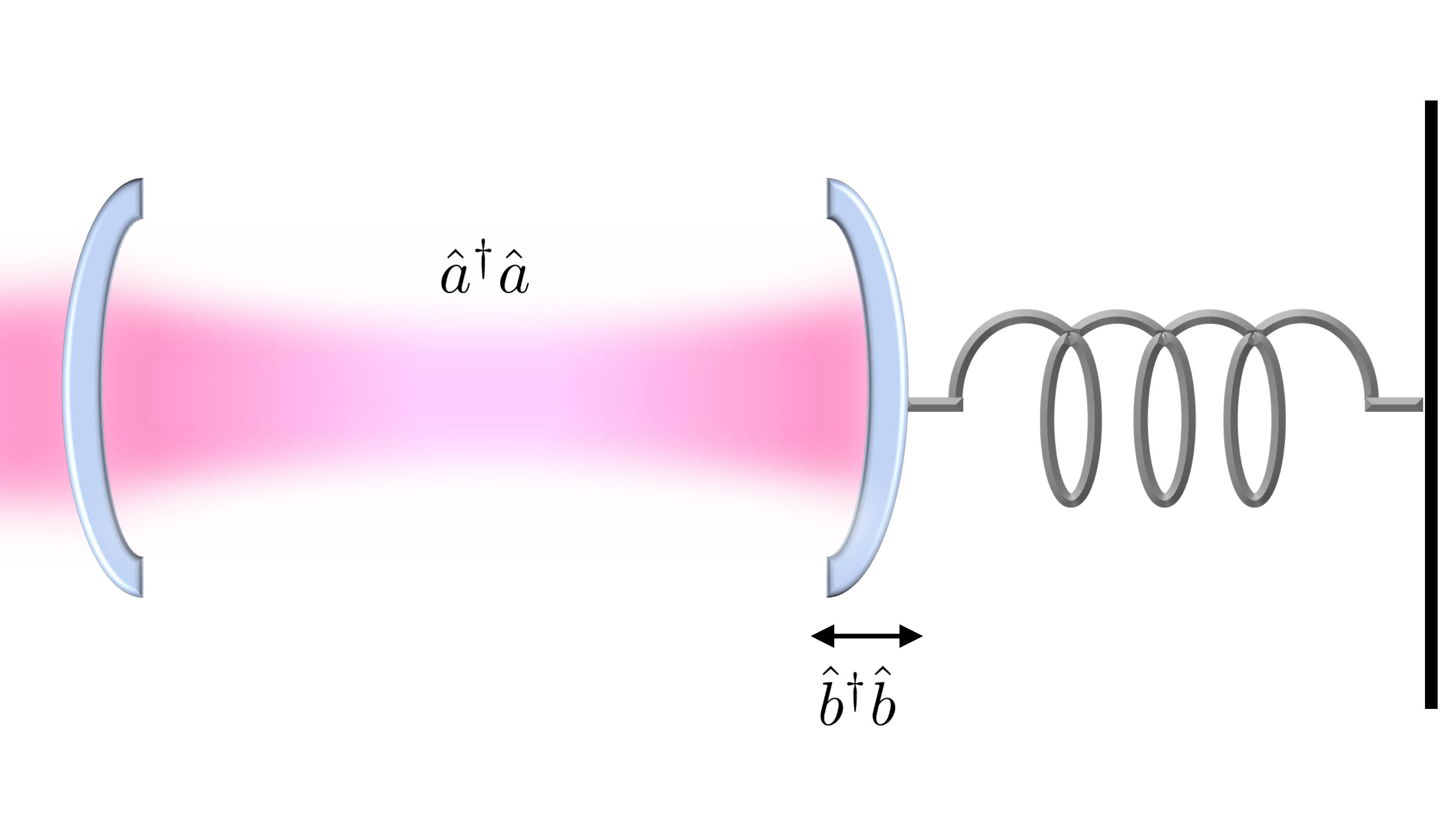}
\caption{Cavity optomechanics is one realization of the Hamiltonian~\eqref{main:time:independent:Hamiltonian:to:decouple}. A semitransparent mirror allows the electromagnetic field to enter the cavity and interact with a moving-end mirror, which therefore affects the frequency of the fundamental modes that can be trapped in the cavity~\cite{bowenbook}. The degree of freedom of the mirror (i.e. its position) can be modeled as a harmonic oscillator coherently interacting with the field.}\label{figure} 
\end{figure}

In this paper we address this problem by computing the ultimate bounds on the estimation of parameters encoded in an optomechanical Hamiltonian with a time-dependent coupling term, a time-dependent mechanical displacement term, and a time-dependent single-mode mechanical squeezing term. 
The time-dependent dynamics of standard optomechanical systems was recently solved~\cite{Bruschi:Xuereb:2018,10.1088/1367-2630/ab1b9e}, and further extended to time-dependent mechanical displacements and squeezing in~\cite{qvarfort2019time}. The methods used to obtain the dynamics have a long history in quantum theory and quantum optics~\cite{Wei_1963,  wilcox1967exponential}.  While for concrete examples we mainly focus on optomechanics, the dynamics we consider (specifically the Hamiltonian ~\eqref{main:time:independent:Hamiltonian:to:decouple}) can be implemented in different setups such as micro-and nanocantilevers, membranes, levitated nanospheres, and optomechanical resonators~\cite{bowenbook,Aspelmeyer:Kippenberg:2014}. 

The paper is organized as follows. We first present the optomechanical Hamiltonian of interest and its analytical solution in Sec.~\ref{sec:system}. We then proceed to define the QFI in Sec.~\ref{sec:metrology} and derive the main result in this work: a general expression for the QFI  of an optomechanical system given the dynamics at hand. Subsequently, in order to demonstrate the applicability of our results, we present three examples of interest: (i) Estimation of the strength of a time-dependent optomechanical coupling (Sec.~\ref{sec:example:1}), (ii) estimation of the strength of a time-dependent linear displacement term (Sec.~\ref{sec:example:2}), and (iii) estimation of the strength of a time-dependent mechanical squeezing term (Sec.~\ref{sec:example:3}). These results are made more concrete in Sec.~\ref{sec:applications}, where we compute the QFI given some example experimental parameters. The paper is concluded by a discussion of our results in Sec.~\ref{sec:discussion}, and some final remarks can be found in Sec.~\ref{sec:conclusions}.

\section{The system}\label{sec:system}
In this section we present the mathematical tools necessary for our work. 
We begin by defining the optomechanical Hamiltonian  and an exact solution of the dynamics. A detailed presentation of the techniques can be found in Appendix~\ref{appendix:decoupling} and the appropriate references mentioned throughout the text. 

\subsection{Optomechanical  Hamiltonian}\label{optomech}
Nonlinear interactions appear in many physical systems, including optomechanical ones, where the bare interaction 
between the electromagnetic field and a mechanical resonator couples the number of photons in the former with the position
of the latter ~\cite{bowenbook,Aspelmeyer:Kippenberg:2014}. An example of an optomechanical system that achieves this nonlinear term is a moving end mirror that forms part of a cavity, which is illustrated in Figure~\ref{figure}.

In this paper we consider the generalized optomechanical Hamiltonian of the form 
\begin{align}\label{main:time:independent:Hamiltonian:to:decouple}
	\hat {H}=  & \hat H_\text{OM} +  \hbar\, \mathcal{D}_1(t) \left(\hat b^\dagger+ \hat b\right)+  \hbar\mathcal{D}_2(t)\left(\hat b^\dagger+ \hat b \right)^2,
\end{align}
where we have introduced the standard optomechanical Hamiltonian $\hat{H}_\text{OM}$ defined by $\hat{H}_\text{OM}:=\hbar\,\omega_\mathrm{c} \hat a^\dagger \hat a + \hbar\,\omega_\mathrm{m} \,\hat b^\dagger \hat b-\hbar\,\mathcal{G}(t) \hat a^\dagger\hat a\,\bigl(\hat b^\dagger+ \hat b\bigr)$, and the (possibly time-dependent) coefficients $\mathcal{G}(t)$, $\mathcal{D}_1(t)$ and $\mathcal{D}_2(t)$.
Here, $\omega_\mathrm{c}$ is the frequency of the light mode with annihilation operator $\hat{a}$, and 
$\omega_\mathrm{m}$ is the trapping frequency of the mechanical mode with annihilation operator $\hat{b}$.

The Hamiltonian~\eqref{main:time:independent:Hamiltonian:to:decouple} reduces to the standard optomechanical Hamiltonian with a constant light--matter coupling when $\mathcal{G}(t) = g_0$, and when $\mathcal{D}_1(t)=\mathcal{D}_2(t)=0$. The time dependence of $\mathcal{G}(t)$ and the additional terms can be obtained in a number of ways: A time-dependent optomechanical coupling is observed in specific experimental systems~\cite{aranas2016split}.
Furthermore, the linear mechanical driving term controlled by $\mathcal{D}_1(t)$ allows for the modeling of an optomechanical system given an externally imposed effect, such as gravitational acceleration~\cite{Qvarfort:Serafini:2018,armata2017quantum}, 
while the single-mode mechanical squeezing term  controlled by $\mathcal{D}_2(t)$ can be obtained by modulating the mechanical frequency~\cite{rashid16,bothner2019cavity}. 

In  what follows, it will be convenient to adopt the dimensionless time $\tau:=\omega_\mathrm{m}\,t$, the dimensionless optical frequency $\Omega_{\rm{c}}:=\omega_\mathrm{c}/\omega_\mathrm{m}$, and the dimensionless Hamiltonian coefficients $\tilde{\mathcal{G}}(\tau):=\mathcal{G}(t)/\omega_\mathrm{m}$, $\tilde{\mathcal{D}}_1(\tau):=\mathcal{D}_1(t)/\omega_\mathrm{m}$, and $\tilde{\mathcal{D}}_2(\tau):=\mathcal{D}_2(t)/\omega_\mathrm{m}$. This means that we will use the rescaled Hamiltonian
\begin{align}\label{main:time:independent:Hamiltonian:to:decouple:dimensionless}
	\hat {H}/(\hbar \, \omega_{\rm{m}})=&\hat{\tilde{H}}_\text{OM}+  \tilde{\mathcal{D}}_1(\tau) \left(\hat b^\dagger+ \hat b\right)+ \tilde{\mathcal{D}}_2(\tau)\left(\hat b^\dagger+ \hat b \right)^2 \, , 
\end{align}
 to compute the dynamics in the following section, and throughout this paper, where $\hat{\tilde{H}}_{\rm{OM}} = \Omega_{\rm{c}} \hat a^\dag\hat a + \hat b^\dag \hat b - \tilde{\mathcal{G}}(\tau) \hat a^\dag \hat a \bigl( \hat b^\dag + \hat b\bigr)$.

\subsection{Decoupling of the time-evolution operator of a nonlinear time-dependent optomechanical Hamiltonian}\label{sub:algebra:techniques}
The main aim of this paper is to provide bounds on precision measurements of parameters that appear in the Hamiltonian~\eqref{main:time:independent:Hamiltonian:to:decouple}. We assume that the parameter of interest can enter into any of the coefficients or frequencies of~\eqref{main:time:independent:Hamiltonian:to:decouple}. 
Therefore, it is necessary to obtain the full time evolution of the system. The time evolution operator corresponding to the Hamiltonian~\eqref{main:time:independent:Hamiltonian:to:decouple} may be expressed as the time-ordered exponential $\hat{U}(\tau):=\overset{\leftarrow}{\mathcal{T}}\exp\bigl[-\frac{i}{\hbar}\int_0^\tau\,d\tau'\,\hat{H}(\tau')\bigr]$. However, this expression is usually cumbersome  to manipulate and only perturbatively applicable. In order to reduce the complexity of the problem, we exploit Lie algebra methods to obtain tractable expressions for the time evolution of the full quantum system~\cite{Wei_1963,wilcox1967exponential}. More specifically, in a first step, we identify the minimal Lie algebra  that generates the time evolution operator. If the minimal Lie algebra is finite, the time evolution operator can be written  in terms of a \textit{finite} product of exponentials of real scalar functions $F_n(\tau)$ multiplied by base elements $\hat{h}_n$ of the Lie algebra, i.e.,~$\hat{U}(\tau) = \prod_n \exp\bigl[ -i  F_n(\tau)\,\hat{h}_n\bigr]$, where the number of factors is equal to the dimension of the  Lie algebra~\cite{Wei_1963}. The scalar functions $F_n(\tau)$ have to be found by solving a set of coupled ordinary differential equations~\cite{Bruschi:Xuereb:2018}.

The time evolution induced by the Hamiltonian~\eqref{main:time:independent:Hamiltonian:to:decouple:dimensionless} has been already decoupled explicitly  using the following set of Hermitian operators as generators of the minimal Lie algebra~\cite{Bruschi:Xuereb:2018,qvarfort2019time}:
\begin{align} \label{eq:lie:algebra}
	\hat N_a^2 &:= (\hat a^\dag \hat a)^2  \nonumber \\
	\hat{N}_a &:= \hat a^\dagger \hat a   &
	\hat{N}_b &:= \hat b^\dagger \hat b \nonumber \\
	\hat{B}_+ &:=  \hat b^\dagger +\hat b &
	\hat{B}_- &:= i\,(\hat b^\dagger -\hat b)  \nonumber\\
	\hat{B}^{(2)}_+ &:= \hat b^{\dagger2}+\hat b^2 &
	\hat{B}^{(2)}_- &:= i\,(\hat b^{\dagger2}-\hat b^2) \nonumber \\
		\hat N_a \, \hat B_+ &:= \hat a^\dag \hat a \, \bigl( \hat b^\dag + \hat b \bigr) & 
	\hat N_a\, \hat B_- &:= i \, \hat a^\dag \hat a \, \bigl( \hat b^\dag - \hat b\bigr).
\end{align} 
It follows that the time evolution operator can be written in the following form 
 \begin{align} \label{eq:explicit:U}
\hat U(\tau)= &  \, e^{-i  J_b \hat{N}_b}e^{-i   J_+ \hat{B}_+^{(2)} } e^{- i  J_- \hat{B}_-^{(2)} } e^{-i(\Omega_{\rm{c}} \tau + F_{{\hat N}_a} )\hat{N}_a} \nonumber \\
\quad&\times e^{- i  F_{\hat N^2_a} \hat N_a^2}  e^{-i (F_{\hat B_+} + F_{\hat N_a \, \hat B_+} \hat{N}_a) \hat{B}_+} \nonumber \\
\quad&\times e^{-i (F_{\hat B_-} + F_{\hat N_a \, \hat B_-} \hat{N}_a) \hat{B}_-} ,
\end{align}
where the explicit forms of the $F$ and $J$ coefficients depend on the functions $\tilde{\mathcal{G}}(\tau)$, $\tilde{\mathcal{D}}_1(\tau)$, and $\tilde{\mathcal{D}}_2(\tau)$ in~\eqref{main:time:independent:Hamiltonian:to:decouple:dimensionless}. Their expressions can be found in Appendix~\ref{appendix:decoupling} and Appendix~\ref{appendix:mechanical} respectively.

By defining the operators $\mathcal{\hat{F}}_\pm :=  F_{\hat B_\pm} + F_{\hat N_a  \, \hat B_\pm} \hat{N}_a$ and $\mathcal{\hat{F}}_{\hat N_a} :=  F_{{\hat N}_a} + F_{{\hat N}^2_a} \,\hat{N}_a$ and using the definition of the Weyl displacement operator $\hat D_b(\beta) = \mathrm{exp}\bigl[ \beta \, \hat b^\dag - \beta^*  \, \hat b \bigr]$, we can rewrite the time evolution operator as\footnote{As $\hat{N}_a$ commutes with all operators in~\eqref{U}, $\hat{N_a}$ and  $\hat N_a^2$ can be treated as $c$-number-valued functions in all manipulations of the exponentials in $\hat{U}(\tau)$. In particular, exponential terms containing only $\hat{N_a}$ and $\hat N_a^2$ and the identity can be freely combined and shifted in $\hat{U}(\tau)$.}
 \begin{align}\label{U}
\hat U(\tau):=& \,\hat {\tilde{U}}_{\rm{sq}}\, e^{-i(\Omega_{\rm{c}} \,\tau + \mathcal{\hat{F}}_{\hat N_a})\hat{N}_a - i \mathcal{\hat{F}}_+ \mathcal{\hat{F}}_- }\, \hat{D}_b(\mathcal{\hat{F}}_- - i\mathcal{\hat{F}}_+) \, ,
\end{align}
where we used the standard formula for the composition of two displacement operators and defined the operator
\begin{align} 
\label{eq:usq} \hat{\tilde U}_{\rm{sq}} = &  \,  e^{-i \, J_b \hat{N}_b}\,\hat{S}_b(2 \, i \, J_+)\,\hat{S}_b(-2 \, J_-), 
\end{align}
using the definition of the squeezing operator $\hat{S}_b(\zeta):=\exp(\frac{1}{2}(-\zeta \hat{b}^{\dagger 2} + \zeta^* \hat{b}^2))$. As already mentioned, the coefficients $J_b$ and $J_\pm$ can be determined by solving a set of differential equations the derivation of which we show in Appendix~\ref{appendix:mechanical}. 

 The form of $\hat U(\tau)$ in~\eqref{U} can now be interpreted as follows: The mechanical oscillator experiences a photon-number dependent displacement through $\hat D_b(\hat{\mathcal{F}_-}-  i \, \hat{\mathcal{F}_+})$, followed by two squeezing operations $\hat S_b(2 \, i \, J_+)$ and $\hat S_b(- 2 \, J_-)$, and a rotation $e^{ -  i \, J_b \, \hat N_b}$. The cavity field is rotated through $e^{ - i (\Omega_{\rm{c}} + F_{\hat N_a}) \hat N_a}$ and then strongly translated by a nonlinear Kerr self-interaction term: $e^{- i \, F_{\hat N_a^2} \hat N_a^2}$. Using a general composition law for squeezing operators given in Appendix~\ref{app:comp}, the full time evolution operator can be reordered and interpreted as subsequent photon number dependent squeezing, displacement and rotation. Details can be found in Appendix~\ref{app:int}.

\subsection{Initial state of the system}

In this paper, we assume that the mechanical element is initially in a thermal state $\hat \rho_\mathrm{Mech.}(T)$ (a standard assumption in the usual regimes of operation), and the light is in a coherent state $\ket{\mu_\textrm{c}}$ (accessible through laser driving). Explicitly, the initial state of the system is
\begin{equation}\label{initial:state}
\hat \rho(0) = \ket{\mu_\textrm{c}}\bra{\mu_\textrm{c}}\otimes \sum_{n=0}^\infty \frac{\tanh^{2n}r_T}{\cosh^2 r_T}\ket{n}\bra{n}\;,
\end{equation}
where $\hat{a}\ket{\mu_\textrm{c}}=\mu_\textrm{c}\ket{\mu_\textrm{c}}$, and where the parameter $r_T$ is defined through the relation   $ r_T=\tanh^{-1} \bigl(\exp[-\frac{\hbar\,\omega_\textrm{m}}{2\,k_\textrm{B}\,T}]\bigr)$,  for which $k_{\rm{B}}$ is Boltzmann's constant and $T$ is the temperature.

\section{Quantum metrology} \label{sec:metrology}
Quantum metrology provides the tools to compute ultimate bounds on precision measurements of parameters contained in a quantum
channel~\cite{paris09}. The general scheme requires an input state
$\hat \rho(0)$, a channel  
that propagates the state, and  $\hat \rho(\theta):=\hat P_\theta\,\hat
  \rho(0)$ with propagator $\hat P_\theta$, 
and depends on a
classical parameter $\theta$ that will be estimated, and a set of
measurements on the final state $\hat\rho(\theta)$. 
The quantum Fisher information (QFI)
$\mathcal{I}_\theta$ allows for the computation of ultimate 
bounds on sensitivity imposed by the laws of
physics~\cite{helstrombook,holevobook}. The QFI is a dimensionful
information measure the inverse of which provides a lower bound to the
variance $\rm{Var}(\theta)$ of 
an unbiased estimator of a parameter $\theta$ through the
quantum Cram\'er--Rao bound (QCRB)
$\mathrm{Var}(\theta)\geq(M \, \mathcal{I}_\theta)^{-1}$~\cite{braunstein1994statistical,braunstein1996generalized,paris09}. The
QCRB is optimized over all possible positive operator-valued
measure  measurements~\cite{peres2006quantum} and all possible
unbiased estimator functions. Its importance arises from the fact that
it can be saturated in the limit of a large number $\mathcal{M}$ 
of measurements, 
independently of how the parameter is  encoded into the state. The optimal measurement is given by a projective measurement onto the eigenstates of the symmetric logarithmic derivative~\cite{braunstein1994statistical}.  Maximum likelihood estimation allows for optimal parameter estimation based on the measurement results, and the maximum likelihood estimator saturates the Cram\'er-Rao bound and becomes unbiased in the limit $\mathcal{M}\rightarrow \infty$. When taken as a measurement prescription rather than a benchmark for the minimal uncertainty of estimating the parameter at the position of its actual value, one faces the problem that the actual value is \textit{a priori} unknown. An adaptive strategy can  be used in this case, where the first measurements provide a rough estimate of the parameter.  The optimal measurement can then be implemented based on this estimate~\cite{barndorff2000fisher}, and iteratively refined.
The QCRB hence constitutes an important benchmark for the ultimate sensitivity that can be achieved (at least in principle when all technical noise problems are solved), and only the fundamental uncertainties due to the quantum state itself remain. 

For unitary channels
that imprint the parameter $\theta$ on an initial state
$\hat \rho(0) = \sum_n \lambda_n \ket{\lambda_n}\bra{\lambda_n}$
according to
{ $\hat\rho(\theta)=\hat U_\theta \hat\rho(0) \hat U_\theta^\dagger$}, 
the quantum Fisher
information can in general be written in the 
form~\cite{pang2014,jing2014}
\begin{align}\label{definition:of:QFI}
\mathcal{I}_\theta
=& \;4\sum_n \lambda_n\,\left(\bra{\lambda_n}\mathcal{\hat H}_\theta^2\ket{\lambda_n} - \bra{\lambda_n}\mathcal{\hat H}_\theta\ket{\lambda_n}^2\right)\nonumber\\
&-8\sum_{n\neq m}
\frac{\lambda_n \lambda_m}{\lambda_n+\lambda_m}
\bigl| \bra{\lambda_n}\mathcal{\hat H}_\theta \ket{\lambda_m}\bigr|^2,
\end{align}
where the second sum is over all terms with $\lambda_n+\lambda_m\ne
0$, $\lambda_n$ is the eigenvalue of the eigenstate $\ket{\lambda_n}$,
and the Hermitian operator $\mathcal{\hat H}_\theta$ is defined by
$\mathcal{\hat H}_\theta=-i\hat U^\dagger_\theta \partial_\theta{\hat
  U}_\theta$~\cite{pang2014,jing2014}. 
The expression~\eqref{definition:of:QFI} was derived for the so-called phase shift Hamiltonian, where the dependence of $\hat U_\theta=\exp\bigl[ -  i \hat H(\theta)\bigr]$ on $\theta$ is through an arbitrary (differentiable) $\hat H(\theta)$.  While we here consider single-parameter estimation, it should in principle be possible to extend these methods to multi-parameter metrology. However, this is beyond the scope of this paper.

In this paper, the channel ${\hat U}_\theta$ is the time evolution operator~\eqref{U}, and the parameter $\theta$ to be estimated is chosen depending on the specific case of interest. Using the decoupled time evolution operator~\eqref{U}, we find
\begin{equation}\label{intermediate:qfi}
	\mathcal{\hat H}_\theta \,= \, \mathcal{\hat H}_{\hat N_a} + \sum_{s\in\{+,-\}} \hat{\mathcal{H}}_s \hat{B}_s + E\hat{N}_b + F\hat{B}_+^{(2)} + G \hat{B}_-^{(2)}\,,
\end{equation}
with $\mathcal{\hat H}_{\hat  N_a} = A \hat{N}_a^2 + B \hat{N}_a + K$,
where $K$ is a constant, and $\hat{\mathcal{H}}_\pm  = C_\pm + C_{\hat
  N_a,\pm} \hat{N}_a$.
The $c$-valued functions $A$, $B$, $C_+$, $C_{\hat N_a,+}$, $C_-$, and
$C_{\hat N_a,-}$ are { given in~\eqref{qfi:coefficients} in Appendix~\ref{appendix:sensing}.}

The QFI ~\eqref{definition:of:QFI} can now be computed by taking the
expectation values of the operator-valued terms
in~\eqref{intermediate:qfi} with respect to the initial state
$\hat{\rho}(0)$ (see~\eqref{initial:state}). The eigenvectors
$\ket{\lambda_n}$ and eigenvalues $\lambda_n$
in~\eqref{definition:of:QFI} are given by $\ket{\lambda_n} =
\ket{\mu_\textrm{c}}\otimes \ket{n}$ and
$\lambda_n=\tanh^{2n}(r_T)/\cosh^2(r_T)$ for the initial state~\eqref{initial:state}. 
This leads us to the main result of this paper, which is an expression for the quantum Fisher information for general metrology with the nonlinear optomechanical Hamiltonian~\eqref{main:time:independent:Hamiltonian:to:decouple}: 
\begin{align} \label{eq:main:result:QFI}
\mathcal{I}_\theta = 4 \biggl[& (4 |\mu_\textrm{c}|^6+6 |\mu_\textrm{c}|^4+|\mu_\textrm{c}|^2)A^2
+  2(2 |\mu_\textrm{c}|^4+|\mu_\textrm{c}|^2)AB\nonumber\\
&  +|\mu_\textrm{c}|^2 B^2  +\cosh(2 \, r_T) \sum_{s\in\{+,-\}} C^2_{\hat N_a,s}  |\mu_c|^2 \nonumber \\
&+\frac{1}{\cosh(2 \, r_T)} \sum_{s\in\{+,-\}} (C_s + C_{\hat N_a,s} |\mu_c|^2)^2 \nonumber\\
&  +  4\frac{\cosh ^2(2 r_T)}{\cosh ^2(2 r_T)+1} \left( F^2 + G^2 \right)   \biggr] \, .
\end{align}
A detailed derivation of ~\eqref{eq:main:result:QFI} is given in
Appendix~\ref{appendix:sensing}. The explicit form of the functions
$A$, $B$, $C_\pm$, $C_{\hat N_a,\pm}$, $F$, and $G$ 
depends on the parameter $\theta$ that we wish to estimate. They also
contain the time dependence of $\hat U_\theta$. 

Let us briefly comment on the form~\eqref{intermediate:qfi} of the QFI. The full explicit expression~\eqref{eq:main:result:QFI} is not particularly revealing, since the coefficients can take different forms depending on the dynamics at hand and the estimation parameter of interest. We note that, in general, the system scales strongly with the parameter $|\mu_{\rm{c}}|$, in particular with the leading term $16 \, |\mu_{\rm{c}}|^6 A^2$. It arises from the fact that $\hat{\mathcal{H}}_\theta$ contains the term $\hat N_a^2$, which when squared yields an expectation value~\eqref{app:eq:Na:overlaps} containing terms of order $|\mu_{\rm{c}}|^8$ and $|\mu_{\rm{c}}|^6$. The eight-order terms cancel, while the leading behavior of $|\mu_{\rm{c}}|^6$ is retained.  

We also note that the term multiplying the first sum in~\eqref{eq:main:result:QFI} scales exponentially with the temperature parameter $r_T$. This implies that, in certain cases, the QFI will increase with the temperature  parameter $r_T$ of the initial thermal state. 
Such a behavior is reminiscent of the increase of QFI with temperature for the measurement of frequency of a simple harmonic oscillator~\cite{braun_ultimate_2011}, which in turn can be attributed to the increasing sensitivity of higher excited Fock states of the resonator.
 For estimating the frequency of the mechanical oscillator or the
cavity, it should be mentioned that in principle also the operators
$\hat a,\hat a^\dagger$ depend on $\omega_{\mathrm{c}}$ (and correspondingly
$\hat b,\hat b^\dagger$ depend on $\omega_{\mathrm{m}}$).  This can be seen most easily
from the fact that the Fock states, i.e.,~the eigenstates of $\hat
a^\dagger \hat a$, depend on $\omega_{\mathrm{c}}$ via the oscillator length,
which becomes clear when writing them in position basis. This dependence
becomes important 
for times much smaller than the period (see~\cite{braun_ultimate_2011}, and for a careful analysis of frequency
estimation of a harmonic oscillator see~\cite{binder_quantum_2019}).  Neglecting
this contribution means that the QFI for frequency estimation is
underestimated.  In what follows, we focus on estimation of parameters
other than frequency, however, where this plays no role.

\begin{figure*}[t!]
\subfloat[ \label{fig:QFI:g0:time}]{%
  \includegraphics[width=0.337\linewidth, trim = 0mm 0mm 0mm -2mm]{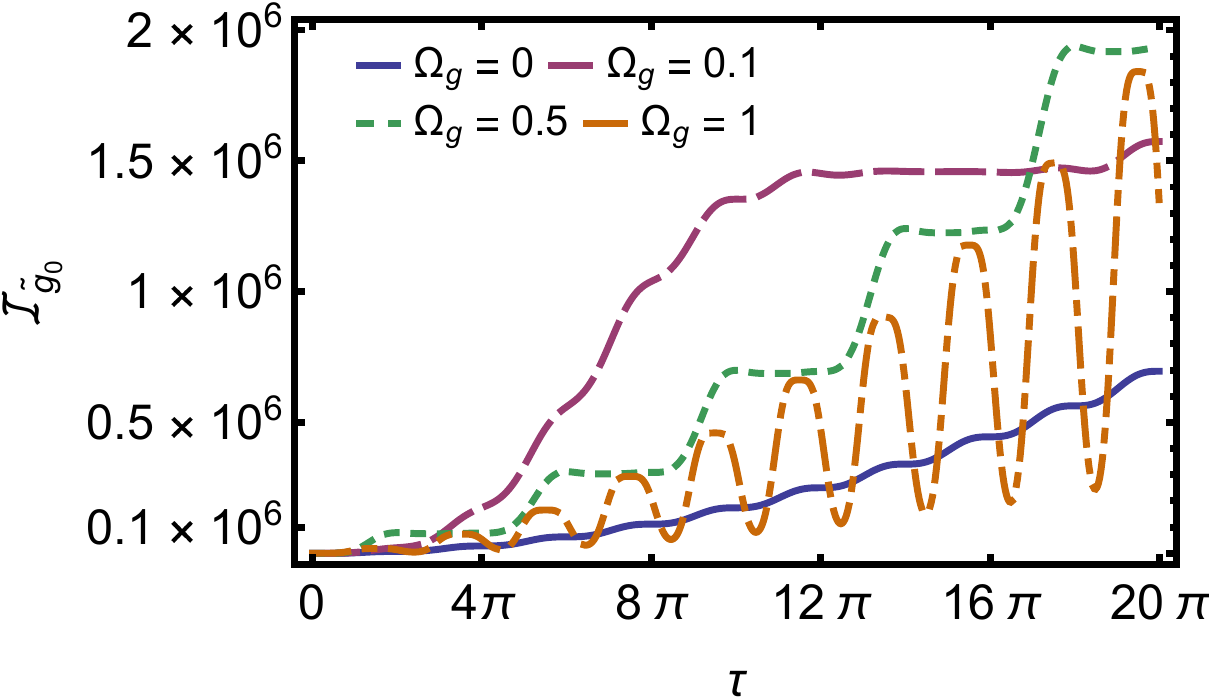}%
} \hspace{1cm}
\subfloat[ \label{fig:QFI:d1:time}]{%
  \includegraphics[width=0.31\linewidth, trim = 0mm 0mm 0mm 2mm]{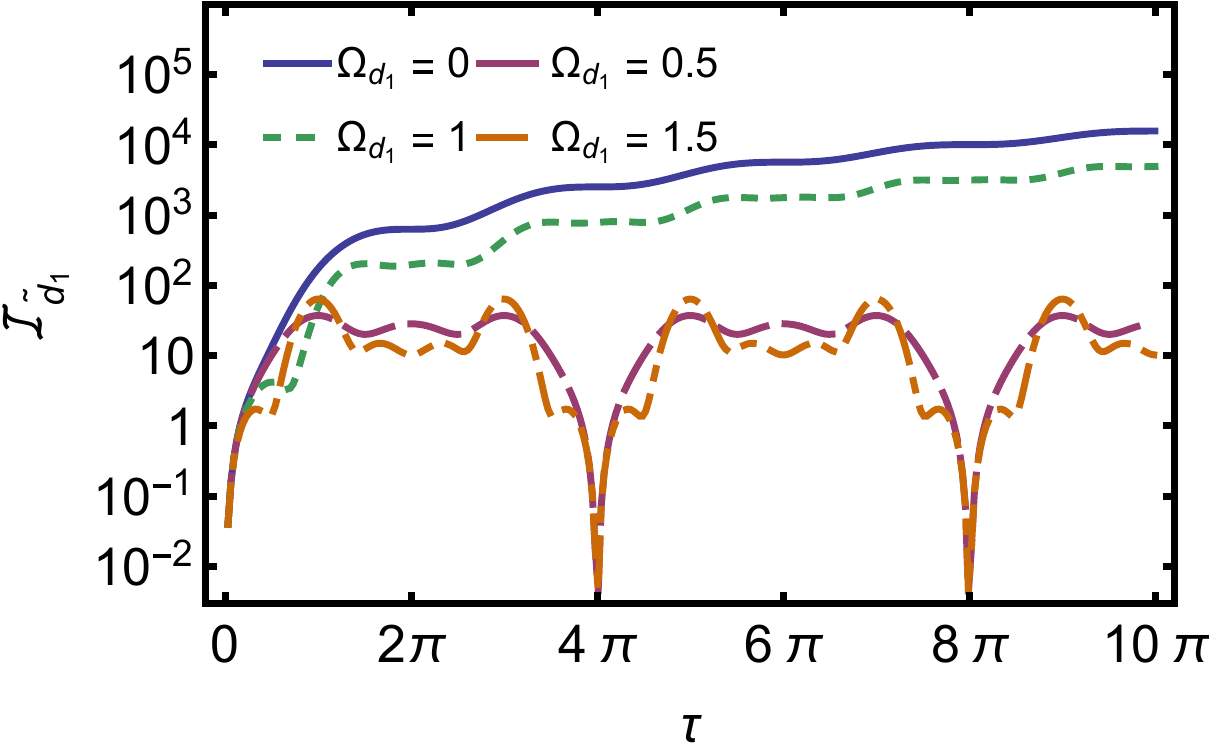}%
}\hfill
\caption{QFI   for estimation of $\textbf{(a)}$ $\tilde{g}_0$ and \textbf{(b)} $\tilde{d}_1$ as a function of dimensionless time $\tau$ for different values of $\Omega_{g}$ (respectively, $\Omega_{\tilde{d}_1}$). While overall the QFI tends to increase with time in both cases, modulations with the period of the harmonic oscillators are clearly visible. For $\Omega_{\tilde{d}_1}\ge 1$, $\mathcal I_{\tilde{d}_1}$ is bounded from above, and a doubling of the period is observed. For a discussion of the relation of the QFI plotted here to the Heisenberg limit, see the Discussion ( Sec.~\ref{sec:discussion}).
}  
\label{fig:QFI:time}
\end{figure*}

\section{Examples}\label{sec:examples}
In this section, we demonstrate the applicability of the main result~\eqref{eq:main:result:QFI} by considering  three concrete scenarios: (i) estimation of the strength of a time-dependent optomechanical coupling, (ii) estimation of the strength of a time-dependent linear mechanical displacement, and (iii) estimation of a time-dependent mechanical squeezing term.

\subsection{Example (i): Estimating the strength of an oscillating optomechanical coupling $\tilde{\mathcal{G}}(\tau)$} \label{sec:example:1}

\begin{figure*}[t!]
\subfloat[ \label{fig:QFI:g0:Omegag:sweep}]{%
  \includegraphics[width=0.3\linewidth, trim = 0mm 0mm 2mm 0mm]{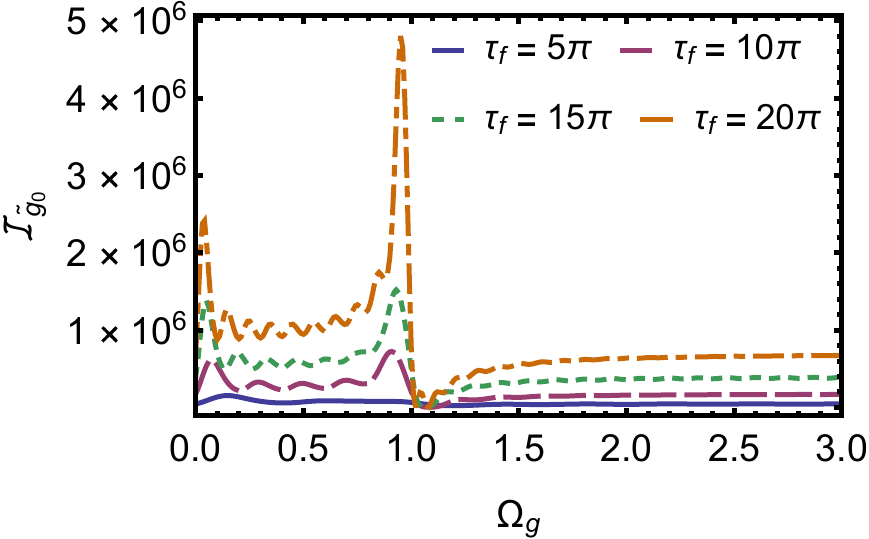}%
} \hspace{1cm}
\subfloat[ \label{fig:QFI:d1:Omegad1:sweep}]{%
  \includegraphics[width=0.31\linewidth, trim = 0mm 0mm 0mm 0mm]{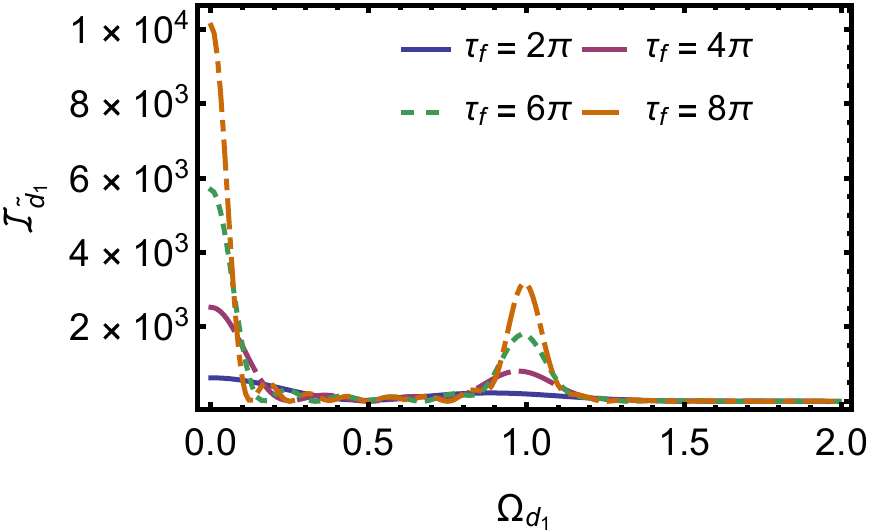}%
}\hfill
\caption{QFI for estimation of  \textbf{(a)} $\tilde{g}_0$  and  \textbf{(b)} $\tilde{d}_1$ for different frequencies. Parameters are $\tilde{g}_0 = 1$, $\epsilon = 0.5$ and $\mu_{\rm{c}} = 1$ for \textbf{(a)}, and 
$\tilde{g}_0 = 1$ and $\mu_{\rm{c}} = 1$ for \textbf{(b)}.
We find that the constant case and the resonances perform best. While the resonance at about $\Omega_g = 1$ gives the best QFI for the estimation of $\tilde{g}_0$, the QFI for the estimation of $\tilde{d}_1$ at the resonance $\Omega_{d_1} = 1$ is smaller compared with when $\tilde{\mathcal{D}}_1(\tau)$ is constant. }
\label{fig:QFI:g0}
\end{figure*}

Characterizing the nonlinear coupling in optomechanical systems is a key task when calibrating an experimental system. The case of a constant coupling $\tilde{\mathcal{G}}(\tau) \equiv \tilde{g}_0$ has already been thoroughly considered~\cite{bernad2018optimal}. As an example application of our methods we therefore compute the QFI for estimating the strength $\tilde{g}_0$ of an \textit{oscillating} optomechanical coupling $\tilde{\mathcal{G}}(\tau)$. We assume that it has the functional form 
\begin{equation} \label{eq:oscillating:light:matter:coupling}
\tilde{\mathcal{G}} (\tau) := \tilde{g}_0 \left( 1 + \epsilon \sin{(\Omega_g \tau)} \right)\;,
\end{equation}
where $\tilde{g}_0 = g_0/\omega_{\rm{m}}$ is the strength of the coupling, $\epsilon$ is the oscillation amplitude, and  $\Omega_g = \omega_g/\omega_{\rm{m}}$. We additionally assume that $\tilde{\mathcal{D}}_1=\tilde{\mathcal{D}}_2=0$. 

A nonlinear coupling of this form appears for levitated 
microscopical particles such as microspheres or nanospheres in Paul traps, where the time-dependent modulation is caused by micromotion of the sphere~\cite{millen2015cavity,fonseca2016nonlinear,aranas2016split}.   A time-varying coupling is also essential for the purpose of exploiting and exploring the quantum thermodynamics of optomechanical systems~\cite{brunelli2015out}, and the same dynamics can be simulated by an optomechanical system where the photon number couples quadratically to the mechanics with $\hat a^\dag \hat a \hat x_{\rm{m}}^2$~\cite{Bruschi:Xuereb:2018}. 

Using the form ~\eqref{eq:oscillating:light:matter:coupling} of the coupling we can compute the $F$ coefficients in~\eqref{sub:algebra:decoupling:solution} explicitly. First of all, we find that whenever $\tilde{\mathcal{D}}_1 (\tau) = 0$, it follows that $F_{\hat N_a} = F_{\hat B_+} = F_{\hat B_-} = 0$, and when $\tilde{\mathcal{D}}_2 = 0$, we have $J_b =\tau$ and $ J_\pm =0$. Then, the remaining \textit{non-zero} coefficients in~\eqref{eq:main:result:QFI} are given by 
\begin{align}\label{c1}
A &= -\partial_\theta F_{\hat N_a^2}-2 F_{\hat N_a \, \hat B_-}\partial_\theta F_{\hat N_a\, \hat B_+}\;,\\\nonumber
 C_{\hat N_a,\pm} &= -\,\partial_\theta F_{\hat N_a \, \hat B_\pm} \;.
\end{align}
The QFI thus becomes
\begin{align} \label{eq:QFI:g0}
\mathcal{I}_{\tilde{g}_0}= & \, 4 \, |\mu_c|^2\,  \biggl(  
\left( 4 \, |\mu_{\rm{c}}|^4 + 6  \, |\mu_{\rm{c}}|^2 +1 \right) A^2
 \nonumber \\
&+  \cosh(2 r_T)  \, \left(1+ \frac{|\mu_{\rm{c}}|^2}{\cosh^2{(2 r_T)}} \right)  \sum_{s\in\{+,-\}} C_{\hat N_a,s}^2 \biggr). 
\end{align}
We observe that the QFI increases for increasing temperatures, 
which is due to the higher occupied phonon states (see the discussion after~\eqref{eq:main:result:QFI}). 
The remaining coefficients $A$ and $C_{\hat N_a, \pm}$ in~\eqref{eq:QFI:g0} need to be complemented with the appropriate expressions~\eqref{sub:algebra:decoupling:solution} for the nonzero $F$ coefficients.  To compute them, we note that $\xi=e^{-i\tau}$ in our case (see~\eqref{app:eq:def:of:xi} and the expressions for the $F$ coefficients in Sec.~\ref{app:coeff:time:dependent:c1}).
The resulting expression for the QFI~\eqref{eq:QFI:g0} is long and cumbersome, so we display it
in~\eqref{app:QFI:eq:g0:general:Omega} in
Appendix~\ref{app:coeff}.

We plot $\mathcal{I}_{\tilde{g}_0}$~\eqref{app:QFI:eq:g0:general:Omega} as a function of time $\tau$ for various frequencies $\Omega_g$ in Figure~\ref{fig:QFI:g0:time}. We note that the different choices of $\Omega_g$ lead to distinct oscillation patterns in $\mathcal{I}_{\tilde{g}_0}$.
Furthermore, we plot $\mathcal{I}_{\tilde{g}_0}$ as a function of $\Omega_{g}$ in Figure~\ref{fig:QFI:g0:Omegag:sweep} for the values $\tilde{g}_0 = |\mu_{\rm{c}}| = 1$, and $r_T = 0$. We note that the QFI peaks  at the resonance frequency $\Omega_g = 1$, but only at later times $\tau \gg 1$.  At earlier time, the peak occurs for values of $\Omega_g \leq 1$. 

When the coupling modulation occurs at mechanical resonance with $\Omega_g \rightarrow 1$, the QFI takes on a more compact form. We present the full expression in~\eqref{eq:app:QFI:c1:res} in Appendix~\ref{app:coeff}.  We can simplify it even further by noting that, at large time-scales $\tau \gg 1$, the first term of~\eqref{eq:app:QFI:c1:res} dominates. Furthermore, when the
mechanical oscillator in the vacuum state with $r_T = 0$,  and when
the optomechanical coupling is much greater than the oscillation
amplitude, $\tilde{g}_0 \gg \epsilon$, and when $\epsilon \ll 1$, the expression simplifies 
significantly to 
\begin{align}\label{eq:QFI:g0:asymptotic}
\mathcal{I}_{\tilde{g}_0}^{(\rm{res, app})} \sim& 16 \, \tilde{g}_0^2 \, \tau^2\, |\mu_{\rm{c}}|^2 \left( 4  \, |\mu_{\rm{c}}|^4 + 6 \, |\mu_{\rm{c}}|^2 + 1 \right) \nonumber \\
&\quad\quad\quad\quad\quad\quad\quad\quad\quad\times\left( 1 - \epsilon \sin( \tau) \right) \, , 
\end{align}
where we kept terms up to $\epsilon$. 
As expected, when $|\mu_{\rm{c}}|^2 $ is zero (no initial cavity mode excitations) or $\tilde{g}_0$ is zero  (no coupling), the QFI vanishes. The same can be seen from the full expression~\eqref{eq:app:QFI:c1:res}. 

The expression~\eqref{eq:QFI:g0:asymptotic} shows that the leading time-dependence of the QFI is quadratic.  This is also true for the more general nonresonant case (see~\eqref{app:QFI:eq:g0:general:Omega}). However, in both cases there are important time-dependent modulations that can lead to a rather large gain or loss of QFI in relatively short time (see e.g.~$\Omega_g=1$ in Fig.~\ref{fig:QFI:g0:time}), which makes the choice of time of measurement crucial.  
\subsection{Example (ii): Estimating a parameter in the linear displacement $\tilde{\mathcal{D}}_1(\tau)$} \label{sec:example:2}
The case of constant $\tilde{\mathcal{D}}_1$ has already been explored  in the context of gravimetry~\cite{Qvarfort:Serafini:2018, armata2017quantum}. Here we extended the analysis by the case of a time-dependent driving $\tilde{\mathcal{D}}_1(\tau)$,  which leads to a signal that is generally easier to detect experimentally compared with a static signal.

We consider a periodic modulation of the mechanical driving term $\tilde{\mathcal{D}}_1(\tau)$ of the form 
\begin{equation} \label{eq:d1:time:dependent}
\tilde{\mathcal{D}}_1 (\tau) = \tilde{d}_1 \, \cos(\Omega_{d_1} \tau) \;,
\end{equation}
where $\tilde{d}_1$ is the dimensionless driving strength and $\Omega_{d_1} = \omega_{d_1}/\omega_{\rm{m}}$ is the oscillation frequency of the driving. A coupling of this form can, for example, be produced in levitating setups by applying any  ac  electric field to the system~\cite{rashid17} that exerts a periodic force to the levitated object. 

We are interested in estimating the driving strength $\tilde{d}_1$ of the time-dependent coupling.  As opposed to the last section,  here we assume that the light--matter coupling is constant with $\tilde{\mathcal{G}}(\tau) \equiv \tilde{g}_0$, and we also assume that $\tilde{\mathcal{D}}_2=0$. This implies that   $\partial_\theta F_{\hat N_a^2} = \partial_\theta F_{\hat N_a \, \hat B_\pm}=0$.  Furthermore, since $\tilde{\mathcal{D}}_2=0$, it follows that $ J_b = \tau$ and $J_\pm = 0$, as well as $\xi(\tau) = e^{- i \, \tau}$. As a result, the following coefficients are zero: $A =C_{\hat N_a,+} = C_{\hat N_a,-}  = F = G = 0$ and the only \textit{nonzero} coefficients that appear in the expression~\eqref{eq:main:result:QFI} of the QFI are 
\begin{align}\label{d1}
 B&=  - \partial_\theta F_{\hat N_a}-2
F_{\hat N_a \, \hat B_-}\partial_\theta F_{\hat B_+}
 \;,\\\nonumber 
C_\pm &= -\partial_\theta F_{\hat B_\pm}\;.
\end{align}
This implies that the QFI for the estimation of $\tilde d_1$ reduces to the expression
\begin{align}\label{qfi:estimate:d1}
\mathcal{I}_{\tilde{d}_1} = 4\,B^2\,|\mu_\textrm{c}|^2 +  \frac{4}{\cosh(2r_T)} \sum_{s\in\{+,-\}} C_s^2\;.
\end{align}
We note that the term $4 \, B^2 \, |\mu_{\rm{c}}|^2$ specifically encodes the nonlinearity; that is, when  $\tilde{g}_0 = 0$ it follows that $B = 0$.

The $F$ coefficients in~\eqref{sub:algebra:decoupling:solution} can now be analytically derived~\eqref{eq:app:F:coefficients:d1}. An explicit expression for $\mathcal{I}_{\tilde{d}_1}$  for general $\Omega_{d_1}$ is given in~\eqref{result:formula:oscillating:d1}. 
 For a constant linear displacement, $\Omega_{d_1}=0$, the $F$ coefficients~\eqref{eq:app:F:coefficients:d1} simplify,  and the QFI takes the simpler expression: 
\begin{align}\label{eq:QFI:d1:general}
\mathcal{I}_{\tilde{d}_1}^{\rm{(const)}} =& 16 \biggl(\tilde{g}_0^2 \left| \mu_c \right| ^2 (\tau-\sin (\tau))^2 +\frac{\sin ^2\left(\tau/2\right)}{\cosh(2 \, r_T)} \biggr) .
\end{align}
The first contribution in this expression originates from the cavity field and its interaction with the  mechanical oscillator, while the second contribution originates from the mechanical oscillator only, which includes the dependence on the temperature  through $r_T$. The origin of the terms can be inferred from the following observation: When either the optical state is the vacuum state (defined by $|\mu_{\rm{c}}| = 0$),
 or the optomechanical coupling is zero (that is, $\tilde{g}_0 = 0$), the contributions from  $B$ vanishes, while the coefficients $C_\pm$ remain nonzero. This situation corresponds to estimating the displacement of a single mechanical element without the cavity.  We note that, in this setting, the enhancement from $|\mu_{\rm{c}}|^2$ is lost, which means that the QFI is reduced overall. We also note that the result in Eq.~\eqref{eq:QFI:d1:general} extends previous findings~\cite{Qvarfort:Serafini:2018, armata2017quantum}  from coherent states to thermal states of the mechanical oscillator. 

When $\tilde{\mathcal{D}}_1(\tau)$ is time dependent~\eqref{eq:d1:time:dependent}, the expression becomes more convoluted~\eqref{result:formula:oscillating:d1}. 
 We plot $\mathcal{I}_{\tilde{d}_1}$ as a function of time $\tau$ for different $\Omega_{d_1}$ in Figure~\ref{fig:QFI:d1:time}. The QFI  continues to increase at large times $\tau$ for the constant ($\Omega_{d_1} = 0$) case and the resonant ($\Omega_{d_1}= 1$) case. For all frequencies $\Omega_{d_1}$ considered, the QFI $\mathcal{I}_{\tilde{d}_1} $ rises very rapidly within about half a period of the mechanical oscillator ($\tau\lesssim \pi$). After the initial rapid increase, the QFI either oscillates or keeps increasing depending on the value of $\Omega_{\tilde{d}_1}$.  Furthermore, in Figure~\ref{fig:QFI:d1:Omegad1:sweep}, we plot $\mathcal{I}_{\tilde{d}_1}$ as a function of the oscillation frequency $\Omega_{d_1}$.  The QFI 
shows a clear local maximum at resonance, where $\Omega_{d_1} = 1$, and another one at $\Omega_{d_1} = 0$, i.e.,~when the displacement $\tilde{\mathcal{D}}_1(\tau) \equiv \tilde{d}_1$ is constant.   

At mechanical resonance  $\Omega_{d_1} = 1$, the expression~\eqref{result:formula:oscillating:d1} simplifies to 
\begin{align} \label{eq:QFI:d1:resonance}
\mathcal{I}_{\tilde{d}_1}^{(\rm{res})} =&  4 \, \tilde{g}_0^2 \, |\mu_{\rm{c}}|^2 \left[\tau+\sin(\tau)\left(\cos(\tau)-2\right)\right]^2   \nonumber\\
& + \frac{\tau^2+2\,\tau\,\sin(\tau)\cos(\tau)+\sin^2(\tau)}{\cosh{(2 r_T)}}. 
\end{align}
We note that 
$\bigl[\tau+\sin(\tau)\left(\cos(\tau)-2\right)\bigr]^2=\bigl[1+\text{sinc}(\tau)\left(\cos(\tau)-2\right)\bigr]^2\,\tau^2$
and
$\tau^2+2\,\tau\,\sin(\tau)\cos(\tau)+\sin^2(\tau)=(1+2\,\text{sinc}(\tau)\cos(\tau)+\text{sinc}^2(\tau))\,\tau^2$,
where $\text{sinc}(x):=\frac{\sin x}{x}$ and
$\text{sinc}(x)\rightarrow 1$ 
for $x\rightarrow0$. This highlights the appearance of terms
proportional to $\tau^2$ in~\eqref{eq:QFI:d1:resonance}. Therefore,
these terms do not oscillate for $\tau \gg 1$ but grow polynomially,
that is, the resonant QFI scales as
$\mathcal{I}_{\tilde{d}_1}^{(\rm{res})} \sim 4 \, \tilde{g}_0^2 \,
|\mu_{\rm{c}}|^2  \, \tau^2 $,  while the QFI for a constant coupling
scales as $\mathcal{I}_{\tilde{d}_1}^{(\rm{const})} \sim 16  \,
\tilde{g}_0^2 \,  |\mu_{\rm{c}}|^2 \,  \tau^2$. All together, this
implies that $\mathcal{I}^{(\rm{const})}_{\tilde{d}_1} \approx  4  \,
\mathcal{I}_{\tilde{d}_1}^{(\rm{res})}$ for $\tau \gg1$.

At higher temperatures, the QFI decreases with larger $r_T$ for both constant~\eqref{eq:QFI:d1:general} and resonant~\eqref{eq:QFI:d1:resonance} displacements. However, the effect differs between the two cases in the $\tau \gg 1$ limit. For $\mathcal{I}_{\tilde{d}_1}^{(\rm{const})}$, the temperature-dependent term is bounded and oscillates with $\tau$, and therefore is completely negligible for $\tau \gg 1$ compared to the term increasing quadratically with $\tau$. For $\mathcal{I}_{\tilde{d}_1}^{(\rm{res})}$, on the other hand, the temperature-dependent term also scales with  $\tau^2$.
  Hence there is resonant buildup of the information contained in the temperature-dependent term, which leads to an advantage for the resonant case when 
  both $r_T$ and  $\tilde{g}_0^2 \,  |\mu_{\rm{c}}|^2 $ are small. 
  The difference between the constant and resonant case is, however, relatively small if $\tilde{g}_0 \gg1$ and $|\mu_{\rm{c}}|^2 \gg 1$, for which the first terms in both~\eqref{eq:QFI:d1:general} and~\eqref{eq:QFI:d1:resonance} dominate and lead to a factor  of 4 in the QFI.

\subsection{Example (iii): Estimating a parameter in the mechanical squeezing $\tilde{\mathcal{D}}_2(\tau)$} \label{sec:example:3}
 In this section, we consider a mechanical squeezing term $\tilde{\mathcal{D}}_2(\tau)$ of the form
\begin{equation} \label{eq:modulated:squeezing}
\tilde{\mathcal{D}}_2(\tau) =  \tilde{d}_2  \,  \cos( \Omega_{d_2} \, \tau) , 
\end{equation}
where $\tilde{d}_2$ is the oscillation amplitude and $\Omega_{d_2}$ is
the frequency.  A modulation of this form can arise
from an external time-dependent shift of the mechanical frequency
$\omega_{\rm{m}}$\footnote{This equivalence is demonstrated explicitly in Appendix D of~\cite{qvarfort2019time}.},  which can be externally imposed employing an oscillating strong optical field or  by applying a pumping voltage in a cantilever setup~\cite{blencowe2004quantum}. Furthermore, a term like this appears as the second-order approximation to a periodic potential, meaning that the inclusion of this term extends our metrology scheme beyond first-order displacements considered in the previous section. In addition, it has previously been shown that modulating the squeezing enhances effects such as entanglement and quantum discord~\cite{farace2012enhancing} -- properties that have previously been found useful for sensing. Lastly, modulating the mechanical squeezing at parametric resonance allows for the creation of increasingly non-Gaussian states~\cite{qvarfort2019time}. 

Our goal is to estimate the
squeezing strength $\tilde{d}_2$ for constant or modulated
couplings.  For simplification we set $\tilde{\mathcal{D}}_1=0$ in this section, 
 and keep $\tilde{\mathcal{G}}(\tau) \equiv \tilde{g}_0$ constant. 
A nonzero mechanical squeezing term affects the full dynamics of the system since it changes the function $\xi(\tau)$~\eqref{app:eq:def:of:xi}, which, in turn, enters into the $F$ coefficients in~\eqref{sub:algebra:decoupling:solution}. The squeezing parameter is   also contained  in the $J$ coefficients, which  may be computed by using the relation~\eqref{app:eq:squeezing:relation}. When $\tilde{\mathcal{D}}_1(\tau) = 0$ we find that  $B = C_\pm = 0$, which means that the general QFI expression~\eqref{eq:main:result:QFI} for estimation of $\tilde{d}_2$ reduces to
\begin{align}\label{eq:QFI:d2}
\mathcal{I}_{\tilde{d}_2} =& 4\, \biggl[  \left( 4\, |\mu_c|^6 + 6\, |\mu_c|^4 + |\mu_c|^2 \right) \, A^2 \nonumber\\
&+ |\mu_c|^2\cosh(2 \, r_T)\left( 1 + \frac{|\mu_c|^2}{\cosh^2(2 \, r_T)} \right) \sum_{s\in\{+,-\}} C^2_{\hat N_a,s}  \nonumber \\
&+  4\frac{\cosh ^2(2 r_T)}{\cosh ^2(2 r_T)+1} \left( F^2 + G^2 \right)\biggr] \, .
\end{align}
When the squeezing term is constant, that is, $\Omega_{d_2} = 0$, the differential equations for the mechanical subsystem evolution~\eqref{differential:equation:written:in:paper} are analytically solvable, as we demonstrate in Appendix ~\ref{app:sec:constant:squeezing}. 
For a time-dependent coupling of the form~\eqref{eq:modulated:squeezing},  however, the mechanical subsystem equations~\eqref{differential:equation:written:in:paper_new} take the form of the Mathieu equation. The Mathieu equation is notoriously difficult to solve numerically, and only has analytic solutions for specific cases. However, it has been shown that perturbative solutions of the form~\eqref{app:eq:RWA} can be obtained at parametric resonance $\Omega_{\tilde{d}_2} = 2$ when $\tilde{d}_2 \ll 1$, i.e., the squeezing strength is small~\cite{qvarfort2019time}. These solutions lead to the same time evolution that can be obtained from the Hamiltonian~\eqref{main:time:independent:Hamiltonian:to:decouple} by employing the rotating wave approximation.

When the squeezing is constant (i.e., $\Omega_{d_2} = 0$), the $F$ coefficients~\eqref{sub:algebra:decoupling:solution}   are given in~\eqref{app:eq:F:coeffs:constant:d2}, and the $J$ coefficients are given in~\eqref{app:eq:J:coeffs:constant:d2}. As a result, the only \textit{nonzero} coefficient of the QFI is
\begin{align}
C_{\hat N_a, +} &= 2 \, \tilde{g}_0 \tau \, ,
\end{align}
which means that the QFI  for estimating a constant squeezing $\tilde{d}_2$ is given by  
\begin{align} \label{eq:app:QFI:d2:constant:app}
\mathcal{I}_{\tilde{d}_2}^{(\rm{const,app})} = 16 \,  \tilde{g}_0^2 \,  \tau^2  \, |\mu_{\rm{c}}|^2 \frac{|\mu_{\rm{c}}|^2+\cosh^2 (2 \, r_T)}{\cosh (2  \, r_T)} ,
\end{align}
where the superscript `app' refers to the fact that our solutions to the dynamics are approximate. 

When the squeezing term is time dependent, with $\tilde{\mathcal{D}}_2(\tau) = \tilde{d}_2 \, \cos(2 \, \tau)$, i.e.,~parametric resonance is assumed,  the $F$ coefficients are given by~\eqref{app:eq:F:coeffs:resonant:d2}, and the $J$ coefficients are given by~\eqref{app:eq:J:coeffs:resonant:d2}. This leads to the following \textit{nonzero} coefficients for the QFI:
\begin{align}
&A = - \tilde{g}_0^2 \tau \,, 
&&C_{\hat N_a, +} =  \tilde{g}_0 \tau \, ,
 &&F = - \tau/2 \, .
\end{align}
The QFI is then given by 
\begin{widetext}
\begin{align} \label{eq:app:QFI:d2:resonance:app}
\mathcal{I}_{\tilde{d}_2}^{(\rm{res,app})} =   \,4 \, \tau^2 \Bigg( & \tilde{g}_0^4 \,  (4\,|\mu_c|^6 + 6\,|\mu_c|^4 + |\mu_c|^2)   + \tilde{g}_0^2\, |\mu_c|^2 \, \frac{|\mu_c|^2 + \cosh^2(2r_T)}{\cosh(2r_T)} + \frac{\cosh^2(2r_T)}{\cosh^2(2r_T)+1} \Bigg)\, . 
\end{align}
\end{widetext}
We note that for the resonant case,  $\mathcal{I}_{\tilde{d}_2}^{(\rm{res,app})}$ scales quadratically with $\tau $ and displays a strong dependence on $\mu_{\rm{c}}$ through the term $|\mu_{\rm{c}}|^6$, while for the constant case, $\mathcal{I}^{(\rm{const,app})}_{\tilde{d}_2}$ only scales with $|\mu_{\rm{c}}|^4$. The QFI for the resonant case also scales with $\tilde{g}_0^4$, which indicates that the strength of the nonlinearity is particularly important for sensing of resonantly modulated squeezing.  Just like in Example (i) in Sec.~\ref{sec:example:1}, we find that the very last term in~\eqref{eq:app:QFI:d2:resonance:app} tends to 1 as $r_T \rightarrow \infty$, but the second-to-last term diverges exponentially as $r_T$ increases, which indicates that a higher temperature $r_T$ contributes positively to the  QFI.  

In the limit $|\mu_{\rm{c}}| \gg 1$, and at zero temperature $r_T = 0$, we find that $\mathcal{I}_{\tilde{d}_2}^{(\rm{const, app})} \sim 16 \, \tilde{g}_0^2 \, \tau^2 \, |\mu_{\rm{c}}|^4$ and $\mathcal{I}_{\tilde{d}_2}^{(\rm{res, app})} = 16 \, \tilde{g}_0^4  \, \tau^2  \,|\mu_{\rm{c}}|^6$, which implies that $\mathcal{I}_{\tilde{d}_2}^{(\rm{res, app})} \sim \tilde{g}_0^2 \, |\mu_{\rm{c}}|^2 \, \mathcal{I}_{\tilde{d}_2}^{(\rm{const, app})}$. It follows that the resonant sensing scheme  might be beneficial for strong light--matter couplings.

\section{ Applications to physical metrology settings}\label{sec:applications}
We have derived a general expression for the QFI  for an optomechanical system operating in the nonlinear regime and discussed three specific examples of parameter estimation scenarios in order to demonstrate how our results can be applied. Our expression can be used to infer the fundamental sensitivity for estimation of any parameters that enter into the Hamiltonian~\eqref{main:time:independent:Hamiltonian:to:decouple}.

To further demonstrate the applicability of these methods, we consider some physical examples of 
parameter values for the following three cases  at resonance:   
estimating the coupling $\tilde{g}_0$ with the exact expression~\eqref{eq:app:QFI:c1:res}, estimating the linear displacement $\tilde{d}_1$~\eqref{eq:QFI:d1:resonance}, and estimating the squeezing parameter $\tilde{d}_2$~\eqref{eq:app:QFI:d2:resonance:app},  which is valid for $\tilde{d}_2 \ll 1$. When we compute the QFI for $\tilde{g}_0$, we set $\tilde{\mathcal{D}}_1 (\tau) = \tilde{\mathcal{D}}_2(\tau) = 0$, and when we compute the QFI for $\tilde{d}_1$ and $\tilde{d}_2$, we keep the optomechanical coupling constant $\tilde{\mathcal{G}}(\tau) \equiv \tilde{g}_0$. In addition, for the estimation of $\tilde{d}_1$ and $\tilde{d}_2$, we set the other coefficient to zero respectively, such that $\tilde{\mathcal{D}}_2(\tau) = 0$ when estimating $\tilde{d}_1$, and $\tilde{\mathcal{D}}_1 = 0$ for estimation of $\tilde{d}_2$. 

The parameters used for all cases include the coupling strength
$\tilde{g}_0 = 10^2$, which can be readily achieved with levitated
systems~\cite{millen2019optomechanics}, a coherent-state parameter of
$|\mu_{\rm{c}}|^2= 10^6$, a temperature of $200$ nK, and a mechanical oscillation frequency $f_{\rm{m}} =10^2$ Hz (which implies the angular frequency  $\omega_{\rm{m}} = 2 \pi \times 10^2\,\rm{rad\, s^{-1}}$). These parameters result in a temperature parameter $r_T = 2.56$. We consider a single
measurement performed at the final time $\tau_f= 2\pi$. The results
can be found in Table~\ref{tab:Values}, where dimensions can be
restored where required by multiplication with the appropriate number of $\omega_{\rm{m}}$.

We now discuss all three cases in detail, where we relate the dimensionless values in Table~\ref{tab:Values} to three physical settings. In all examples, we list our results with three significant digits, however they should be seen as merely indicative of the order of magnitude of the fundamental measurement limit. 

\begin{itemize}
\item[(i)] \textit{Estimation of the amplitude
$\tilde{g}_0$}. The constant case has already been thoroughly explored~\cite{bernad2018optimal}. We therefore focus on a time-dependent coupling at mechanical resonance. We set the oscillation amplitude to $\epsilon =0.5$, and by using $\tilde{g}_0 = 10^2$ and $|\mu_{\rm{c}}|^2 = 10^6$, we find   from~\eqref{eq:app:QFI:c1:res} that the dimensionless QFI becomes $\mathcal{I}_{\tilde{g}_0}^{(\rm{res})} = 3.02 \times 10^{25}$. This implies a single-shot sensitivity of $\Delta\tilde{g}_0 = 1/(\mathcal{I}_{\tilde{g}_0}^{(\rm{res})})^{\frac{1}{2}}  =
1.82\times 10^{-13}$ and a relative sensitivity of $\Delta \tilde{g}_0/\tilde{g}_0 = 1.82 \times 10^{-15}$.           
	\item[(ii)] \textit{Estimation of $\tilde{d}_1$}. The constant case has already been previously considered~\cite{Qvarfort:Serafini:2018,armata2017quantum}. For the resonant case, we find from~\eqref{eq:QFI:d1:resonance} that $\mathcal{I}_{\tilde{g}_0}^{(\mathrm{res})} =1.58 \times 10^{12}$, which implies a single-shot sensitivity of $\Delta \tilde{d}_1 = 7.96\times10^{-7}$. Since  we set $\tilde{d}_1  = 1$ in our example, the relative sensitivity $\Delta \tilde{d}_1/\tilde{d}_1$ takes the same value.  This example can be made more concrete in the context of force sensing. We consider detection of a spatially constant force, which physically corresponds to the system subjected to a linear potential with oscillating slope, which causes the mechanical element to become displaced. Let $\tilde{\mathcal{D}}_1(\tau) = a(\tau) \sqrt{ m/(2 \, \hbar\, \omega_{\rm{m}}^3)}$, where $m$ is the mass of the system, and $a(\tau) = a_0 \, \cos(\Omega_{a} \, \tau)$ is a time-dependent acceleration. We then obtain $\tilde{d}_1= a_0 \, \sqrt{ m/(2 \, \hbar \, \omega_{\rm{m}}^3)}$, in analogy with Example (ii) in Sec.~\ref{sec:example:2}. Since we now are interested in estimating $a_0$ rather than $\tilde{d}_1$, we note that $\partial_{a_0} = \partial_{a_0} \tilde{d}_1 \partial_{\tilde{d}_1}$, and hence the (dimensionful) QFI, becomes $\mathcal{I}_{a_0}^{(\rm{res})} = ( \partial_{a_0} \, \tilde{d}_1)^2 \,\mathcal{I}_{\tilde{d}_1}^{(\rm{res})}$.  To compute a value for the QFI, we consider a   levitated 
 object with a mass $m = 10^{-14}$ kg with an angular  oscillation frequency of $\omega_{\rm{m}} = 2 \pi \times 10^2$ rad $s^{-1}$. Given these values together with the parameters $\tilde{g}_0 = 10^2$, $|\mu_{\mathrm{c}}|^2 = 10^6$, and $T = 200$ nK, which implies $r_T =2.56$,  we find  the dimensionless QFI to be $\mathcal{I}_{\tilde{d}_1}^{(\rm{res})} = 1.58\times 10^{12}$, which after restoring dimensions yields $\mathcal{I}_{a_0}^{(\rm{res})} = 7.48\times 10^{23} \, \rm{m}^{-2}\rm{s}^4$. The sensitivity becomes $\Delta a_0 = 1.16\times 10^{-13}$ ms$^{-2}$, which in turn should allow for measurements of resonant forces of amplitude $m \, \Delta a_0 = 1.16\times 10^{-27}$ N.  
	\item[(iii)] \textit{Estimation of a
constant shift 
or parametric modulation of the cavity frequency $\delta
\omega_{\rm{m}}$}. 
This measurement task corresponds to
Example (iii)  considered in Sec.~\ref{sec:example:3} with $\mathcal{D}_2(t) = \delta\omega_{\rm{m}}(t)$.  We start by assuming a constant squeezing
with $\delta \omega_{\rm{m}}(t) \equiv \delta \omega_{\rm{m}}$. This yields the following dimensionless parameter  $\tilde{d}_2 =
\delta \omega_{\rm{m}}/\omega_{\rm{m}}$, where we chose small values of $\delta \omega_{\rm{m}} /\omega_{\rm{m}}= 0.1$ to ensure the validity of our approximation. Similarly to the above, we are here interested in estimating $\delta \omega_{\rm{m}}$ rather than $\tilde{d}_2$, and we note that $\partial_{\delta \omega_{\rm{m}}} = \partial_{\delta \omega_{\rm{m}}} \tilde{d}_2  \, \partial_{\tilde{d}_2} =(\omega_{\rm{m}})^{-1} \partial_{\tilde{d}_2}$. The  dimensionful QFI therefore becomes $\mathcal{I}_{\delta \omega_{\rm{m}}}^{(\rm{const, app})} = (\omega_{\rm{m}})^{-2} \,\mathcal{I}_{\tilde{d}_2}^{(\rm{const, app})}$. 
Then, we set $\tilde{g}_0 = 10^2$, $|\mu_{\rm{c}}|^2 = 10^6$, and $\omega_{\rm{m}} = 2\pi \times 10^2 \,\rm{rad\, s^{-1}}$, which implies $\delta \omega_{\rm{m}} = 2\pi \times 10\,\rm{rad\, s^{-1}}$, and a temperature of $200$ nK, which yields $r_T = 2.56$. We then find  from~\eqref{eq:app:QFI:d2:constant:app} that 
$\mathcal{I}_{\delta \omega_{\rm{m}}}^{(\mathrm{const, app})} = 1.93\times10^{11}\,\rm{s^{2}\,rad^{-2}}$ ,
which implies a sensitivity  to static shifts of the frequency of $\Delta (\delta \omega_{\rm{m}}) =2.27\times 10^{-6}\,\rm{rad\, s^{-1}}$,  and a relative sensitivity of $\Delta (\delta \omega_{\rm{m}})/\delta \omega_{\rm{m}} = 3.62\times10^{-8}$. Next, we consider the case where the frequency change is time dependent with $\delta \omega_{\rm{m}}(t) = \delta \omega_{\rm{m}}\cos( \omega_0 \, t)$, where the driving is resonant with $\omega_0/\omega_{\rm{m}} = 2$. 
We use the same values as above to find from~\eqref{eq:app:QFI:d2:resonance:app} that the dimensionless QFI is $\mathcal{I}_{\tilde{d}_2}^{(\rm{res,app})} = 6.32\times 10^{28}$, which yields $\mathcal{I}_{\delta \omega_{\rm{m}}}^{( \rm{res, app})} =
1.60 \times 10^{23} \,\rm{s^{2}\,rad^{-2}}$. This implies a sensitivity  to modulated frequency shifts of $\Delta (\delta
\omega_{\rm{m}}) = 2.50 \times 10^{-12} \,\rm{rad\, s^{-1}}$  and a relative sensitivity of 
$\Delta (\delta \omega_{\rm{m}})/\delta \omega_{\rm{m}} = 3.98 \times
10^{-14}$.  
\end{itemize}

{\renewcommand{\arraystretch}{1.2} 
\begin{table}[h!]
\begin{ruledtabular}
\begin{tabular}{Sl Sc Sc} 
 \textbf{Parameter} & \textbf{Symbol}  &  \textbf{Value} \\ 
\hline 
Time of measurement & $\tau_f = \omega_{\rm{m}} \, t $ & $2\pi$ \\
Optomechanical coupling & $ \tilde{g}_0 = g_0/\omega_{\rm{m}}$ & $10^2$   \\
Coherent state parameter & $|\mu_{\rm{c}}|^2$ & $10^6$ \\
Mechanical oscillation frequency & $\omega_{\rm{m}}$ & $2 \pi \times 10^2$ rad s$^{-1}$\\
Thermal state temperature & $T$ & $200$ nK \\
Thermal state parameter & $r_T$ & 2.56 \\ \hline
\multicolumn{3}{Sc}{\textbf{Estimation of $\tilde{g}_0$}}\\ \hline
Amplitude of coupling oscillation & $\epsilon $ & $0.5$   \\ 
\textbf{QFI for estimation of $\tilde{g}_0$}~\eqref{eq:app:QFI:c1:res} & $\mathcal{I}^{(\rm{res})}_{\tilde{g}_0} $ & $3.02\times 10^{25}$ \\ \hline
\multicolumn{3}{Sc}{\textbf{Estimation of $\tilde{d}_1$}}\\ \hline
Linear displacement & $\tilde{d}_1 = d_1/\omega_{\rm{m}}$ & 1 \\
\textbf{QFI for estimation of $\tilde{d}_1$}~\eqref{eq:QFI:d1:resonance}& $\mathcal{I}^{(\rm{res})}_{\tilde{d}_1} $ & $1.58 \times 10^{12}$\\ \hline
\multicolumn{3}{Sc}{\textbf{Estimation of $\tilde{d}_2$}}\\ \hline
Squeezing parameter & $\tilde{d}_2 = d_2/\omega_{\rm{m}} $ & 0.1 \\
\textbf{QFI for estimation of $\tilde{d}_2$}~\eqref{eq:app:QFI:d2:resonance:app} & $\mathcal{I}^{(\rm{res, app})}_{\tilde{d}_2} $ & $6.32\times 10^{28}$\\
\end{tabular} 
\end{ruledtabular}
\caption{The single-shot QFI for estimating the optomechanical coupling strength $\tilde{g}_0$, a linear mechanical displacement strength $\tilde{d}_1$, and a mechanical squeezing strength $\tilde{d}_2$ (all on resonance). In each scheme, we set the other couplings to zero or, in the case of the coupling $\tilde{g}_0$, to a constant. Estimation of $\tilde{g}_0$ and, in certain schemes, $\tilde{d}_2$ corresponds to an internal characterization of the system, while estimation of $\tilde{d}_1$ and $\tilde{d}_2$ yields the sensitivity of the optomechanical system to an external force and its changing amplitude, respectively.  The numbers shown here are dimensionless and can correspond to a variety of physical settings.  Dimensions can be restored when necessary by considering the specific nature of the Hamiltonian couplings, and we provide three such examples in Section~\ref{sec:applications}. For a discussion of how these numbers compare to the Heisenberg limit, see the Discussion section (Sec.~\ref{sec:discussion}).}
\label{tab:Values}
\end{table}

\section{Discussion}\label{sec:discussion}
In the previous sections, we showed how to use solutions of the time evolution induced by the Hamiltonian~\eqref{main:time:independent:Hamiltonian:to:decouple} to obtain bounds on the sensitivity with which some relevant experimental parameters contained in the Hamiltonian can be measured. We gave three explicit examples, however we note that our methods can be extended to a number of additional parameters. Here, we discuss our results and elaborate on a number
of properties of the QFI.

\subsection{The Heisenberg limit}
The Heisenberg limit 
is often associated with a scaling of the sensitivity of a system as
$N^{-1}$ (as opposed to $N^{-1/2}$ for classical systems), where $N$
is the number of physical probes in the system.  However, it should be
kept in mind that this result is derived under rather specific
conditions~\cite{braun_quantum-enhanced_2018}: $N$ distinguishable, noninteracting subsystems, finite-dimensional
Hilbert spaces, and parameter encoding via a unitary evolution with a 
parameter-dependent Hamiltonian~\cite{Giovannetti04,giovannetti2006quantum}.   By coincidence, the $1/N$ (respectively
$1/\sqrt{N}$) scaling is also the scaling of the sensitivity with the average
number of photons with which the phase shift in a Mach--Zehnder
interferometer can be measured. This scaling occurs when a NOON state (respectively, the
coherent state) is used, even though the photons are indistinguishable
bosons with infinite-dimensional Hilbert space, and the photon number
is in both cases only defined on average.  This result follows
immediately from the general expression of the pure state QFI in terms
of the variance of the generator $\hat G$ that generates the unitary
{ transformation} $\hat U_\alpha$ which encodes the parameter $\alpha$ according to
$\hat U_\alpha=e^{i \alpha \hat G}$, together with the phase shift
Hamiltonian $\hat H=\alpha \hat a^\dagger \hat a$. It is, however,
also well known that the scaling with $N$ can be faster than $1/N$ for
the 
estimation of an interaction parameter~\cite{luis_nonlinear_2004,braun_heisenberg-limited_2011},
and this advantage can extend in certain parameter regimes to the
estimation of other parameters 
of an interacting system~\cite{fraisse_coherent_2015} if one has access to the full system. 
In light of the $1/N$ scaling that is often associated with the
Heisenberg limit, our main result~\eqref{eq:main:result:QFI} appears
to indicate scaling beyond the Heisenberg limit due to the term
$|\mu_{\rm{c}}|^6$, which can be written in terms of the initial
average number $N_{\rm{ph}}$ of photons as
$|\mu_{\rm{c}}|^6=N_{\rm{ph}}^3$. A similar scaling has been predicted for the
phase sensitivity of nonlinear optical
systems~\cite{boixo2007generalized}. The $N^3$ term corresponds to a sensitivity
that scales $\propto N_{\rm{ph}}^{-3/2}$, i.e.,~decays faster than the
``Heisenberg limit'' $1/N$. The origin of the
$|\mu_{\rm{c}}|^6$ 
term is clearly the $(\hat a^\dagger \hat a)^2$ term in  $\hat{\mathcal{H}}_{N_a}$ (see~\eqref{intermediate:qfi}).  If one restricts the maximum amount of
energy available, its contribution to the QFI is maximized when the light and mechanics form the  aforementioned NOON
state~\cite{braun_quantum-enhanced_2018}, but $N_{\rm{ph}}$ is replaced by $N_{\rm{ph}}^2$,  i.e.,~the true Heisenberg-limit in
the sense of the smallest possible uncertainty is 
now a $1/N_{\rm{ph}}^2$ scaling of the sensitivity, whereas the coherent state
gives the $1/N_{\rm{ph}}^{3/2}$ found above.  Given that a NOON state is extremely difficult to prepare, especially for highly excited Fock states, the scaling obtained for the coherent state is quite favorable, given this consideration.   
Since the corresponding parameter
$F_{\hat N_a}^2$  depends 
not only on the coupling constant $\tilde{\mathcal{G}}_1$
but also on the
 squeezing parameter $\tilde d_2$ relevant for force sensing, we
have here the remarkable situation that the nonlinear interaction between the
two oscillators not only allows enhanced sensitivity for estimating the
interaction (i.e.,~faster than $1/N_{\rm{ph}}$ scaling of the sensitivity,
but which cannot be
compared to the noninteracting case, as the parameter  $\tilde{g}_0$ does not exist
there), but also, significantly, enables 
enhanced sensitivity of a parameter of the original noninteracting
system! This is a fundamental insight that was possible only through
the exact decoupling scheme used here, and it should be highly useful
for metrology.  In principle one could 
envisage other systems leading to even higher powers of $N_{\rm{ph}}$, if the
Lie algebra of generators in 
$\hat H$ closed after more iterations. We note, however, that the sensitivity to linear displacements  with this system scales as $1/N_{\rm{ph}}^{1/2}$, 
i.e.,~up to a change of prefactor the same sensitivity as for measuring a phase shift with a coherent state.  However, it should be kept in mind that it is the excitation of the optical cavity that determines the sensitivity with which the shift of the mechanical oscillator is measured, and which can be much larger than the initial thermal excitation of the mechanical oscillator.

\subsection{Resonance}
Here we discuss the implications of driving the system at mechanical resonance. The resonance behavior differs for all three examples considered in Sec.~\ref{sec:examples}, which implies a rich and
complicated structure of the QFI.  We here provide a brief discussion of some of the main features observed in this paper.   For estimation of $\tilde{g}_0$, it can be seen in   Figure~\ref{fig:QFI:g0:Omegag:sweep}, where we plotted a frequency sweep of the QFI at various times $\tau_f$, that the onset of the increase of QFI is due to the accumulation of the resonant behavior. In fact, Figure~\ref{fig:QFI:g0:Omegag:sweep} demonstrates  that driving on resonance only provides a significant advantage as $\tau \gg 1$. 
  
For estimations of a linear drive $\tilde{d}_1$, we found that a constant coupling performs better than a time-dependent one. This observation is most likely due to our choice to let the weighting function $\tilde{\mathcal{D}}(\tau) = \tilde{d}_1 \cos(\Omega_{d_1}  \tau)$ oscillate around zero rather than a fixed displacement. 

For estimation of $\tilde{d}_2$, our results are only valid 
 close to parametric resonance, which occurs when $\Omega_{d_2} = 2$.  In all cases considered here, in general, we demonstrated that resonances play an important, but not always beneficial, part in enhancing the sensitivity of a system.

 \subsection{Time-dependence}
 In all three examples we considered, the QFI was found to increase essentially quadratically with dimensionless time $\tau$ to leading order  at resonance. 
 Optomechanical systems are among the most massive quantum systems that can be controlled  in the laboratory to date, and while impressively narrow linewidths have recently been demonstrated experimentally with levitated nanoparticles~\cite{pontin2019}, achieving long quantum coherence
 times is still a challenging task.  In the pioneering experiments reported in~\cite{oconnell_quantum_2010} the fitted $T_2$ dephasing time of a nanomechanical oscillator with resonance frequency of 6\,GHz was about 20\,ns, corresponding to a maximally achievable $\tau\simeq 754$. 
Given a finite available measurement time limited by the decoherence time, our results show that the precise timing of the measurements and the choice of frequency ratios are crucial for optimizing the overall sensitivity per square root of hertz.  It is a major benefit of our method that the precise time dependence  of the QFI can be obtained in such a nonlinear and possibly driven or parametrically modulated optomechanical system.

\section{Conclusion}\label{sec:conclusions}
We have derived a general expression for the QFI for a nonlinear optomechanical system with a
time-dependent light--matter coupling term, a time-dependent linear
mechanical displacement term, and a time-dependent single-mode
mechanical squeezing term in the Hamiltonian.  The expression for the
QFI can be used to compute the optimal sensitivity bounds for the
estimation of any 
parameter which enters into any of the terms in the Hamiltonian. Most
importantly, our methods include the treatment of arbitrary
time-dependent effects, which offers significant advantages for
experimental schemes since time-varying signals can be more easily
distinguished from a typical random noise floor than constant ones.  

To demonstrate the applicability of the expression and our methods, we
computed the QFI for three specific examples: (i) estimating the
strength of an oscillating optomechanical coupling, (ii) estimating
the amplitude of an oscillating linear mechanical displacement term, and, (iii)
estimating the amplitude of a resonant time-dependent mechanical
squeezing term. We derived exact and asymptotic expressions for the
QFI in the first two cases, as well as an approximate expression
based on perturbative solutions for a squeezing term modulated at
resonance.  

Our results include a number of interesting phenomena.
Most remarkable is the fact that the nonlinear interaction leads, for
large population of the cavity, to a
drastically increased sensitivity not only for the coupling,  but also the frequency shift of the mechanical oscillator, and hence to the measurement of spatially linearly varying forces.
Secondly, we find that resonances, where the oscillation frequency of the driving matches the mechanical oscillation frequency of the system, or  in case of parametric driving twice the oscillation frequency, can increase the QFI for measuring the coupling or the linear shift (and hence spatially constant forces) substantially. 
Thirdly, we find that the temperature of
the initial mechanical thermal state is not always detrimental for the
sensitivity, and might even sometimes aid estimation of the parameter
in question. More work is needed to establish how this effect can be harnessed for settings that include the potentially detrimental effects of decoherence due to the coupling to a thermal environment, the influence of which on the dynamics was neglected so far.

Finally, while we have analyzed three relevant examples in detail, the methods can be applied to the 
measurement of a large number of
internal and external effects that act on the optomechanical systems,
as long as they can be modeled via the coefficients in the Hamiltonian
we consider. It should be kept in mind, however,  that our results are proofs of existence: they show
that a joint measurement of the cavity and mechanical oscillator
exists that allows one to reach the described sensitivities in the
limit of infinitely { many} measurements.  More
work will be required to understand how the different effects in the
Hamiltonian interact to 
enhance or decrease the sensitivity, and to find physically feasible
measurements that saturate the bounds.   In addition the
question of the effect of decoherence needs to be addressed.
Nevertheless, our results clearly demonstrate the potential of
optomechanical systems, and more generally of 
harmonic oscillators coupled via the radiation-pressure coupling, for 
strongly enhanced sensitivity in the measurement of very small forces.

\section*{Acknowledgments}
We thank Julien Fra\"isse, Doug Plato,  Francesco Alberalli, Antonio Pontin, Nathana\"{e}l Bullier, Peter F. Barker, and Ivette Fuentes 
for useful comments and discussions. 
S.Q. acknowledges support from the Engineering and Physical Sciences Research Council (EPSRC) and the EPSRC Centre for Doctoral Training in Delivering Quantum Technologies and thanks the University of Vienna for its hospitality. D.R. would like to thank the Humboldt Foundation for
supporting his work with their Feodor Lynen Research
Fellowship. This work was supported by the European Union's Horizon 2020 Research and Innovation program under Grant  No.\ 732894 (Future and Emerging Technologies Proactive Hybrid Optomechanical Technologies). D.E.B. thanks the Institute for Quantum Optics and Quantum Information in Vienna and the Eberhard Karls Universit\"at T\"ubingen for their hospitality.

\bibliographystyle{apsrev4-1}
\bibliography{Ref}

\onecolumngrid

\newpage

\appendix

\section{Decoupling time-dependent dynamics}\label{appendix:decoupling}
Here we discuss the basic elements that led to the decoupling of the form~\eqref{U}. All details of the techniques and procedures can be found in~\cite{qvarfort2019time}.
The algebra basis operators are
\begin{align}\label{basis:operator:Lie:algebra}
	\hat{N}_a &:= \hat a^\dagger \hat a & 
	\hat{N}_b &:= \hat b^\dagger \hat b & \hat{N}^2_a &:= (\hat a^\dagger \hat a)^2 \nonumber \\
	\hat{B}_+ &:=  \hat b^\dagger +\hat b &
	\hat{B}_- &:= i\,(\hat b^\dagger -\hat b) &
	 & \nonumber\\
	\hat{B}^{(2)}_+ &:= \hat b^{\dagger2}+\hat b^2 &
	\hat{B}^{(2)}_- &:= i\,(\hat b^{\dagger2}-\hat b^2) &
	 &  \nonumber\\
	\hat{N}_a\,\hat{B}_+ &:= \hat{N}_a\,(\hat b^{\dagger}+\hat b) &
	\hat{N}_a\,\hat{B}_- &:= \hat{N}_a\,i\,(\hat b^{\dagger}-\hat b) \, . &
	 & 
\end{align}
 The time-evolution operator is 
 \begin{align}
\hat U(\tau):=& 
\,\hat{\tilde{U}}_{\rm{sq}}(\tau)\, e^{-i(\Omega_{\rm{c}} \,\tau + \mathcal{\hat{F}}_{\hat N_a})\hat{N}_a}\,  e^{-i\,(F_{\hat B_+} + F_{\hat N_a \, \hat  B_+} \hat{N}_a)\,\hat{B}_+}\,e^{-i\,(F_{\hat B_-} + F_{\hat N_a  \, \hat B_-} \hat{N}_a) \,\hat{B}_-}\,,
\end{align}
where $\hat{\mathcal{F}}_{\hat N_a} = F_{\hat N_a} + F_{\hat N_a^2} \, \hat N_a$, and the expression of $\hat{\tilde{U}}_{\rm{sq}}$ is 
\begin{align}\label{decoupling:form:to:be:used}
\hat{\tilde{U}}_{\rm{sq}} &= \overleftarrow{T} \exp\biggl[ - i \int^\tau_0 \mathrm{d}\tau'\left((1 \,+2 \, \tilde{\mathcal{D}}_2(\tau'))\hat{N}_b+ \tilde{\mathcal{D}}_2(\tau')\hat{B}^{(2)}_+\right) \biggr].
\end{align}
 The action of $\hat{\tilde{U}}_{\rm{sq}}$ on the mode operator $\hat{b}$ is given by $\hat{\tilde{U}}_{\rm{sq}}^\dag\,\hat{b}\,\hat{\tilde{U}}_{\rm{sq}}=\alpha(\tau)\,\hat{b}+\beta(\tau)\,\hat{b}^\dag$.
The Bogoliubov $\alpha(\tau)$ and $\beta(\tau)$ coefficients read
\begin{align} \label{eq:alpha:beta:definition}
\alpha(\tau)=&\frac{1}{2}\,\left[P_{11}(\tau)+P_{22}(\tau)-i\,\int_0^\tau\,d\tau'\,P_{22}(\tau')-i\,\int_0^\tau\,d\tau'\,(1+4\,\tilde{\mathcal{D}}_2(\tau'))\,P_{11}(\tau')\right] \, ,\nonumber\\
\beta(\tau)
  =&\frac{1}{2}\,\left[P_{11}(\tau)-P_{22}(\tau)+i\,\int_0^\tau\,d\tau'\,P_{22}(\tau')-i\,\int_0^\tau\,d\tau'\,(1+4\,\tilde{\mathcal{D}}_2(\tau'))\,P_{11}(\tau')\right] \, ,
\end{align}
and the functions $P_{11}$ and $P_{22}$ can be found by integrating
(with $\dot A\equiv dA(\tau)/d\tau$)
\begin{align}\label{differential:equation:written:in:paper}
\ddot{P}_{11}+(1 +4\,\tilde{\mathcal{D}}_2(\tau))\,P_{11}=& \, 0 \, ,\nonumber\\
\ddot{P}_{22}-\frac{4\,\dot{\tilde{\mathcal{D}}}_2(\tau)}{1 +4\,\tilde{\mathcal{D}}_2(\tau)}\,\dot{P}_{22}+(1+4\,\tilde{\mathcal{D}}_2(\tau))\,P_{22}=&\, 0,
\end{align}
together with the initial conditions $P_{11}(0)=P_{22}(0)=1$ and $\dot{P}_{11}(0)=\dot{P}_{22}(0)=0$.

Let us rewrite the above equations in terms of $P_{11}$ and 
\begin{equation}
	I_{P_{22}} := \int_0^\tau d\tau' P_{22}(\tau')\,.
\end{equation}
Then, the governing differential equations become equivalent, i.e.
\begin{align}\label{differential:equation:written:in:paper_new}
\ddot{P}_{11}+(1 +4\,\tilde{\mathcal{D}}_2(\tau))\,P_{11}= \, & \, 0 \,,  \nonumber\\
\ddot{I}_{P_{22}} + (1 +4\,\tilde{\mathcal{D}}_2(\tau))\,I_{P_{22}}= \, & \, 0 \,,
\end{align}
which can be verified by dividing by $1 +4\,\tilde{\mathcal{D}}_2(\tau)$ and taking the time derivative. The initial conditions for $I_{P_{22}}$ follow from those for $P_{22}$ as $I_{P_{22}}(0) = 0$ and
$\dot {I}_{P_{22}}(0) = 1$.
Furthermore, using the differential equation for $P_{11}$, we find
\begin{align} \label{app:eq:bogoliubov:expressions}
\alpha(\tau) =& \frac{1}{2}\,\left[ P_{11} - i I_{P_{22}} + i\frac{d}{d\tau} ( P_{11}  - i I_{P_{22}} )\right] \, ,\nonumber\\
\beta(\tau) =& \frac{1}{2}\,\left[ P_{11} + i I_{P_{22}} + i\frac{d}{d\tau} ( P_{11}  + i I_{P_{22}} )\right].
\end{align}
Furthermore, we define 
\begin{equation} \label{app:eq:def:of:xi}
	\xi := \alpha + \beta^* = P_{11} - i I_{P_{22}} \,,
\end{equation}
which implies $\alpha = (\xi + i \dot{\xi})/2$ and $\beta = (\xi^* + i \dot{\xi}^*)/2$.

The functions for the decoupling of the time-evolution operator ~\eqref{U} have been computed in~\cite{qvarfort2019time} and we reprint them here  ($\Re\xi$ and $\Im\xi$ denote the real and imaginary part of $\xi$, respectively):
\begin{align}\label{sub:algebra:decoupling:solution}
F_{\hat{N}_a}=& -2 \,\int_0^\tau\,d\tau'\,\tilde{\mathcal{D}}_1(\tau')\,\Im\xi(\tau')\int_0^{\tau'}d\tau''\,\tilde{\mathcal{G}}(\tau'')\,\Re\xi(\tau'') -2  \int^\tau_0\,d\tau' \,\tilde{\mathcal{G}}(\tau')\, \Im \xi(\tau') \, \int^{\tau'}_0 \,d\tau''\, \tilde{\mathcal{D}}_1(\tau'') \, \Re \xi(\tau'') \, , \, \nonumber\\
F_{\hat{N}^2_a}=&  \, 2  \,\int_0^\tau\,d\tau'\,\tilde{\mathcal{G}}(\tau')\,\Im\xi(t')\int_0^{\tau'}d\tau''\,\tilde{\mathcal{G}}(\tau'')\,\Re\xi(\tau'') \, ,\nonumber\\
F_{\hat{B}_+}=& \,\int_0^\tau\,d\tau'\,\tilde{\mathcal{D}}_1(\tau')\,\Re\xi(\tau') \, ,\nonumber\\
F_{\hat{B}_-}=&-  \,\int_0^\tau\,d\tau'\,\tilde{\mathcal{D}}_1(\tau')\,\Im\xi(\tau') \, ,\nonumber\\
F_{\hat{N}_a\,\hat{B}_+}=&- \,\int_0^\tau\,d\tau'\,\tilde{\mathcal{G}}(\tau')\,\Re\xi(\tau') \, ,\nonumber\\
F_{\hat{N}_a\,\hat{B}_-}=& \,\int_0^\tau\,d\tau'\,\tilde{\mathcal{G}}(\tau')\,\Im\xi(\tau') \, .
\end{align}

Finally, two special scenarios give us the following analytical expressions for $\xi$.

\begin{itemize}
	\item[1] For $\tilde{\mathcal{D}}_2(\tau)=0$, we obtain $P_{11}=\cos(\tau)$ and $I_{P_{22}}=\sin(\tau)$, which leads to $\xi= e^{-i\tau}$.
 
	\item[2] When the squeezing term is modulated at frequency $\Omega_{d_2}$ with $\tilde{\mathcal{D}}_2(\tau) = \,\tilde{d}_2\,  \cos(\Omega_{d_2} \, \tau)$, it follows that the solutions to~\eqref{differential:equation:written:in:paper_new} coincide with the solutions to the Mathieu equations. This equation is notoriously difficult to solve, but a set of perturbative solutions were   given in Eq. (E.15) in~\cite{qvarfort2019time}. The solutions are valid for 
$\tilde{d}_2\ll 1$ and $\tau \gg 1$ and yield 
\begin{align} \label{app:eq:RWA}
\xi(\tau) =& \,  e^{- i \, \tau} \, \cosh (\tilde{d}_2 \, \tau) + i \, e^{i \, \tau} \,  \sinh (\tilde{d}_2 \, \tau) \, .
\end{align} 
\end{itemize}

\section{Commutator relations and expectation values }\label{appendix:Fisher:nonlinear:commutators}
In the appendices below, the following expressions must be evaluated by commuting the exponentials through the expression in the middle. We list them and their solutions here for reference. 
\begin{align}\label{app:eq:commutators}
e^{ i x \hat B_- ^{(2)} } \,  \hat B_+^{(2)} \, e^{-i x \hat B_-^{(2)} } =&  \hat B_+^{(2)} \cosh(4 x) +  \left( 2 \, \hat N_b+ 1 \right)  \sinh(4 x) \, ,  \nonumber \\
e^{i x \hat B_+^{(2)}} \hat B_-^{(2)}  e^{- i x \hat B_+^{(2)}} =& \hat B_-^{(2)}  \cosh(4 x) - \left( 2 \, \hat N_b+ 1 \right) \sinh(4 x) \, , \nonumber \\
e^{i x \hat B_-^{(2)} }\, \hat N_b \, e^{- i x \hat B_-^{(2)}} =& \hat N_b \,  \cosh(4x) + \hat B_+^{(2)} \frac{1}{2} \sinh(4x) + \sinh^2(2x) \, \mathds{1} \, , \nonumber \\
e^{i x \hat B_+^{(2)} } \,\hat N_b \, e^{- i x \hat B_+^{(2)}} =& \hat N_b \,  \cosh(4x) -  \hat B_-^{(2)} \frac{1}{2} \sinh(4x) + \sinh^2(2x) \, \mathds{1} \, , \nonumber \\
e^{i\,x\,\hat B_+}\, \hat N_b \,e^{-i\,x\,\hat B_+}=& \, \hat N_b-\hat B_-\,x+x^2\,\mathds{1}\, , \nonumber\\
e^{i\,  x\,\hat B_- }\, \hat N_b \,e^{ -i \, x\,\hat B_-}=&\, \hat N_b+ \hat B_+\,x+x^2\,\mathds{1}\, ,\nonumber\\
e^{i\,x\,\hat B_+ }\,\hat B_+^{(2)} \,e^{-i\,x\,\hat B_+}=&\,  \hat B_+^{(2)} +2\,\hat B_-\,x-2\,x^2\,\mathds{1}\, ,\nonumber\\
e^{i \, x\, \hat  B_-}\,\hat B_+^{(2)} \,e^{  - i \, x\, \hat B_-}=&\,  \hat B_+^{(2)} +2\,\hat B_+ \,x+2\,x^2\,\mathds{1}\, ,\nonumber\\
e^{i\,x\,\hat B_+}\,\hat B_-^{(2)} \,e^{-i\,x\,\hat B_+}=&\, \hat B_-^{(2)} -2\,\hat B_+\,x\, ,\nonumber\\
e^{ i \, x\,\hat B_-}\,\hat B_-^{(2)} \,e^{-i \, x\,\hat B_-}=&\, \hat B_-^{(2)}  + 2\,\hat B_-\,x\, ,\nonumber\\
e^{i\,x\,\hat B_+}\,\hat B_-\,e^{-i\,x\,\hat B_+}=&\, \hat B_--2\,x\,\mathds{1}\, .\nonumber\\
e^{i \, x\,\hat B_-}\,\hat B_+ \,e^{-i \, x\,\hat B_- }=&\, \hat B_+ + 2\,x\,\mathds{1} \, .
\end{align}
Furthermore, we need a number of expectation values in order to compute the QFI. They are 
\begin{align}
\braket{n|\hat B_+^2|n}
&= 2n+1\;,\nonumber \\
\braket{n|\hat B_-^2|n}
&= 2n+1\;,\nonumber\\
\braket{n|(\hat B_+^{(2)})^2|n}
&= 2n^2+2n+2\;,\nonumber\\
\braket{n|(\hat B_-^{(2)})^2|n}
&= 2n^2+2n+2\;,\nonumber\\
\braket{n|\hat B_+\hat B_-|n}
&= i\;,\nonumber\\
\braket{n|
\hat B_-\hat B_+|n}
&= -i\;,\nonumber\\
\braket{n|\hat B_+^{(2)}\hat B_-^{(2)}|n}
&= 2i(2n+1)\;,\nonumber\\
\braket{n|\hat B_-^{(2)}\hat B_+^{(2)}|n}
&= -2i(2n+1)\;,
\end{align}
as well as
\begin{align} \label{app:eq:Na:overlaps}
\braket{\mu_{\rm{c}}|\hat N_a^4|\mu_{\rm{c}}}
&= |\mu_{\rm{c}}|^8+6|\mu_{\rm{c}}|^6+7|\mu_{\rm{c}}|^4+|\mu_{\rm{c}}|^2\;,\nonumber\\
\braket{\mu_{\rm{c}}|\hat N_a^3|\mu_{\rm{c}}}
&= |\mu_{\rm{c}}|^6+3|\mu_{\rm{c}}|^4+|\mu_{\rm{c}}|^2\;,\nonumber\\
\braket{\mu_{\rm{c}}|\hat N_a^2|\mu_{\rm{c}}}
&= |\mu_{\rm{c}}|^2(1+|\mu_{\rm{c}}|^2)\;,
\end{align}
and 
\begin{align}
\braket{n|\hat B_+|m}
&= \sqrt{m+1}\delta_{n,m+1}+\sqrt{m}\delta_{n,m-1}\;,\nonumber\\
\braket{n|\hat B_-|m}
&= i\left(\sqrt{m+1}\delta_{n,m+1}-\sqrt{m}\delta_{n,m-1}\right)\;,\nonumber\\
\braket{n|\hat B_+^{(2)}|m}
&= \sqrt{m+1}\sqrt{m+2}\delta_{n,m+2}+\sqrt{m}\sqrt{m-1}\delta_{n,m-2}\;,\nonumber\\
\braket{n|\hat B_-^{(2)}|m}
&= i \left(\sqrt{m+1}\sqrt{m+2}\delta_{n,m+2} -\sqrt{m}\sqrt{m-1}\delta_{n,m-2}\right)\;.
\end{align}

\section{Treatment of the mechanical squeezing subsystem}\label{appendix:mechanical}

In this Appendix, we  decouple the time evolution of the mechanical subsystem and interpret the time evolution operator in terms of subsequent squeezing, displacement and rotation.

\subsection{Decoupling the mechanical subsystem}
In order to compute the QFI for measurements of parameters in $\tilde{D}_2(\tau)$, we must find an analytic expression for $\hat{\tilde{U}}_{\rm{sq}}(\tau)$. To obtain the coefficients $J_b$ and $J_\pm$, we will follow methods outlined in~\cite{moore2016tuneable, 10.1088/1367-2630/ab1b9e, qvarfort2019time}. 

The operator $\tilde{\hat{U}}_{\rm{sq}}$ is given by 
\begin{align}\label{decoupling:form:to:be:used}
\hat{\tilde{U}}_{\rm{sq}} &= \overleftarrow{T} \exp\biggl[ - i \int^\tau_0 \mathrm{d}\tau'\left(\,(1 \,+2 \, \tilde{\mathcal{D}}_2(\tau'))\hat{N}_b+ \tilde{\mathcal{D}}_2(\tau')\hat{B}^{(2)}_+\right) \biggr].
\end{align}
We want to find an analytic expression in terms of operators that we can treat individually. We make the following ansatz:
\begin{align}\label{app:ansatz}
\hat{\tilde U}_{\rm{sq}}= \exp[-i  \, J_b  \, \hat{N}_b]\,\exp[-i \,  J_+ \hat B_+^{(2)}]\,\exp[ - i  \, J_- \, \hat B_- ^{(2)} ] \, .
\end{align}
We then  differentiate the ansatz with respect to time $\tau$ to obtain 
\begin{align}  \label{app:subsystem:decoupling:differentiated}
\dot{\hat{\tilde{U}}}_{\rm{sq}} \, \hat{\tilde{U}}_{\rm{sq}}^\dag &=  - i \,  \dot{J}_\theta  - i  \, \dot{J}
_+ e^{ - i \, J_b  \, \hat{N}_b } \, \hat B_+^{(2)} \, e^{ i  \, J_b \, \hat{N}_b}  - i \, \dot{J}_- \, e^{ - i \,  J_b \, \hat{N}_b } \,e^{ - i J_+ \hat B_+^{(2)} } \hat B_-^{(2)} \, e^{i J_+ \, \hat B_+^{(2)}} \, e^{i  \, J_b \, \hat{N}_b}\, .
\end{align}
By using the commutator relations~\eqref{app:eq:commutators},~\eqref{app:subsystem:decoupling:differentiated} can be written purely as terms proportional to the operators $\hat N_b$, $\hat B_+^{(2)}$ and $\hat B_-^{(2)}$: 
\begin{align}
\dot{\hat{\tilde{U}}}_{\rm{sq}} \, \hat{\tilde{U}}_{\rm{sq}}^\dag =& \,   - i  \, \dot{J}_\theta \, \hat{N}_b - i \dot{J}_+  \left(   \cos(2 J_b) \hat B_+^{(2)} - \sin (2 J_b) \hat B_-^{(2)} \right)  \nonumber \\
&- i \dot{J}_- \, \left[ \cosh(4\,J_+)\,\left( \cos(2 J_b)  \, \hat B_-^{(2)} + \sin(2 J_b) \hat B_+^{(2)} \right) + 2\,\sinh(4\,J_+)\,\hat{N}_b-4\,J_+\right]\, .
\end{align}
Now we set this equal to the expression under the integral~\eqref{decoupling:form:to:be:used},
\begin{align}
\,(1 \,+2 \, \tilde{\mathcal{D}}_2(\tau))\hat{N}_b+ \tilde{\mathcal{D}}_2(\tau)\hat{B}^{(2)}_+ =& \dot{J}_\theta \, \hat N_b +  \dot{J}_+  \left(   \cos(2 J_b) \hat B_+^{(2)} -  \sin (2 J_b) \hat B_-^{(2)} \right)  \nonumber \\
&+  \dot{J}_- \, \left[ \cosh(4\,J_+)\,\left( \cos(2 J_b)  \, \hat B_-^{(2)} +  \sin(2 J_b) \hat B_+^{(2)} \right) + 2\,\sinh(4\,J_+)\,\hat N_b-4\,J_+ \right] \, .
\end{align}
We then use the linear independence of the operators in order to write down the following differential equations
\begin{align}
(1 \,+2 \, \tilde{\mathcal{D}}_2(\tau)) &=  \dot{J}_b  + 2\,  \, \dot{J}_- \, \sinh(4\,J_+) \, , \nonumber \\
\tilde{\mathcal{D}}_2 ( \tau) &= \dot{J}_+\cos(2 J_b) + \dot{J}_- \, \cosh(4\,J_+)\,\sin(2 J_b) \, ,  \nonumber \\
0 &= - \dot{J}_+ \sin(2 J_b) + \dot{J}_- \, \cosh(4 J_+) \cos( 2 J_b)  \, , 
\end{align}
which can be simplified into the following first-order coupled differential equations: 
\begin{align} \label{eq:diff:equations:Js}
\dot{J}_b &=  1 + 2\, \tilde{\mathcal{D}}_2(\tau) \,  \left( 1 - \sin(2 J_b) \tanh(4 J_+) \right)\,,  \nonumber \\
\dot{J}_+ &=    \tilde{\mathcal{D}}_2(\tau) \,  \cos(2 J_b) \, , \nonumber \\
\dot{J}_-  &= \tilde{\mathcal{D}}_2(\tau) \,  \frac{\sin(2 J_b)}{\cosh(4 J_+)} \, .
\end{align}
These equations do not in general allow for analytic solutions. In the main text, we proceed with estimations of parameters in $\tilde{D}_2(\tau)$ by evaluating these equations numerically.

%
\subsection{The time evolution interpreted}\label{app:int}

Using a general composition law for squeezing operators (see Appendix~\ref{app:comp}), we can write~\eqref{eq:usq} as 
\begin{equation} \label{eq:compact:squeezing}
	\hat{\tilde U}_{\rm{sq}} \dot{=} e^{-i ( J_b +  \varphi_J ) \hat{N}_b }\, \hat{S}_b(\rm{arctanh}( |\zeta_J| )e^{i\arg(\zeta_J)}) \,,  
\end{equation}
where $\dot{=}$ indicates equivalence up to a global phase, and where
\begin{align} \label{eq:squeezing:new:coeffs}
	\varphi_J = & \,  \arctan(\tanh(2J_+)\tanh(2J_-)) \, , \nonumber \\
	\zeta_J = & \, \frac{i\tanh(2J_+) - \tanh(2J_-)}{1-i\tanh(2J_+)\tanh(2J_-)}\, .
\end{align}
With the commutation law for displacement and squeezing,
we obtain 
 \begin{align}\label{eq:intuitiveform}
\hat U(\tau)=& 
\,e^{-i \, (\Omega_{\rm{c}} \,\tau + \mathcal{\hat{F}}_{\hat N_a})\hat{N}_a - i \mathcal{\hat{F}}_+ \mathcal{\hat{F}}_- }\, e^{-i ( J_b + \varphi_J)\hat{N}_b } \hat{D}_b(\hat\gamma)  \hat{S}_b\left(\rm{arctanh}(|\zeta_J|) e^{i\arg(\zeta_J)}\right) \,,
\end{align}
where 
\begin{align}
	\hat\gamma = &  \, \frac{(\mathcal{\hat{F}}_- - i\mathcal{\hat{F}}_+)}{\sqrt{1-|\zeta_J|^2}} - e^{i\arg(\zeta_J)} \frac{(\mathcal{\hat{F}}_- + i\mathcal{\hat{F}}_+)|\zeta_J|}{\sqrt{1-|\zeta_J|^2}} \,.
\end{align}
By rewriting $\hat U(\tau)$ in the form~\eqref{eq:intuitiveform}, we can interpret the time evolution as the following subsequently performed operations: a squeezing, a photon number dependent displacement, and a photon number dependent rotation.

%
\subsection{Derivation of the squeezing composition law}\label{app:comp}
We start from the unitary representation of the squeezing operator 
\begin{equation}
	\hat{U}_\mathrm{sq} = e^{-\frac{r}{2} e^{i\theta} \hat{b}^{\dag 2} + \frac{r}{2} e^{-i\theta} \hat{b}^{2}}
	= e^{-\frac{i}{2} \hat{\mathbb{X}}^\dag \boldsymbol{H}_{\rm{sq}} \hat{\mathbb{X}}}\,,
\end{equation}
which is sometimes also called $\hat S(z)$ where $z$ is a complex number such that $z = r \, e^{i \theta}$, and where we have defined 
\begin{equation}
\boldsymbol{H}_{\rm{sq}} = \left(\begin{array}{cc} 0 & -ir e^{i\theta} \\ ir e^{-i\theta}  & 0 \end{array} \right) \quad \mathrm{and} \quad \hat{\mathbb{X}} = \left(\begin{array}{c} \hat{b} \\ \hat{b}^\dag \end{array} \right)\,.
\end{equation}
The corresponding symplectic representation is given by $S_\mathrm{sq} = e^{\, \boldsymbol{\Omega} \boldsymbol{H}_{\rm{sq}}}$, where the symplectic form in this particular basis is 
\begin{equation}
\boldsymbol{\Omega} = \left(\begin{array}{cc} -i & 0 \\ 0 & i \end{array} \right) \,.
\end{equation}
This leads to the symplectic form of the squeezing operation
\begin{equation}
\boldsymbol{S}_\mathrm{sq}(r,\theta)  = \left(\begin{array}{cc} \cosh(r) & - e^{i\theta} \sinh(r) \\  -  e^{-i\theta} \sinh(r) & \cosh(r) \end{array} \right)\,.
\end{equation}
Therefore, we can write two subsequent squeezing operations as 
\begin{eqnarray}\label{eq:doublesqueeze:first}
\boldsymbol{S}_\mathrm{sq}(r_1,\theta_1) \boldsymbol{S}_\mathrm{sq}(r_2,\theta_2)  & = \begin{pmatrix} S_{11} & S_{12} \\ S_{21} & S_{22} \end{pmatrix} \, , 
\end{eqnarray}
where the matrix elements are given by 
\begin{align} \label{app:eq:squeezing:elements:first}
S_{11} &=\cosh(r_1)\cosh(r_2) + e^{i(\theta_1 - \theta_2)} \sinh(r_1)\sinh(r_2) \, , \nonumber \\
S_{12} &= S_{21}^*=   -\left(e^{i\theta_1} \sinh(r_1)\cosh(r_2) + e^{i\theta_2} \cosh(r_1)\sinh(r_2) \right)  \, , \nonumber \\
S_{22} &= \cosh(r_1)\cosh(r_2) + e^{-i(\theta_1 - \theta_2)} \sinh(r_1)\sinh(r_2) \, .
\end{align}
The unitary representation of a rotation is
\begin{equation}
	\hat{U}_R = e^{-\frac{ia}{2}(\hat{b}^\dag\hat{b} + \hat{b}\hat{b}^\dag)} \, ,
\end{equation}
which corresponds to the symplectic matrix
\begin{equation}
\boldsymbol{S}_{R}(a) = \left(\begin{array}{cc} e^{-ia} & 0 \\ 0 & e^{ia} \end{array} \right)\,.
\end{equation}
A consecutive application of a squeezing and a rotation gives
\begin{equation}\label{eq:rotsqueeze}
\boldsymbol{S}_{R}(a) \boldsymbol{S}_\mathrm{sq}(r_3,\theta_3)  = \left(\begin{array}{cc} e^{-ia}\cosh(r_3) & - e^{i(\theta_3-a)} \sinh(r_3) \\  - e^{-i(\theta_3-a)} \sinh(r_3) & e^{ia}\cosh(r_3) \end{array} \right)\,.
\end{equation}
Identification of the elements in~\eqref{eq:doublesqueeze:first} and~\eqref{eq:rotsqueeze} leads to
\begin{align}
	\cosh(r_3)= & |\cosh(r_1)\cosh(r_2) + e^{i(\theta_1 - \theta_2)} \sinh(r_1)\sinh(r_2)| \, ,\nonumber \\
	\sinh(r_3)= & |\cosh(r_1)\sinh(r_2) + e^{i(\theta_1 - \theta_2)} \sinh(r_1)\cosh(r_2)|\,.
\end{align}
Furthermore, 
\begin{equation}
	e^{i\theta_3} = \frac{\cosh(r_3)}{\sinh(r_3)} \frac{  e^{i\theta_1} \sinh(r_1)\cosh(r_2) + e^{i\theta_2} \cosh(r_1)\sinh(r_2) }  { \cosh(r_1)\cosh(r_2) + e^{i(\theta_1 - \theta_2)} \sinh(r_1)\sinh(r_2) } \, ,
\end{equation}
and, dividing $S_{11}$ by $S_{22}$,
\begin{equation}
	e^{-2ia} = \frac{ \cosh(r_1)\cosh(r_2) + e^{i(\theta_1 - \theta_2)} \sinh(r_1)\sinh(r_2) }{ \cosh(r_1)\cosh(r_2) + e^{-i(\theta_1 - \theta_2)} \sinh(r_1)\sinh(r_2)} \, .
\end{equation}
Defining $t_j = \tanh(r_j) e^{i\theta_j}$, we find
\begin{equation} \label{app:eq:identities:for:tj}
	t_3 = \tanh(r_3) e^{i\theta_3} = \frac{  t_1 + t_2 }  { 1 + t_1 t_2^*} \, ,\quad \mathrm{and} \quad e^{-2ia} = \frac{ 1 + t_1 t_2^*}{ 1 + t_1^* t_2} \, ,
\end{equation}
and the composition law for squeezing operators
\begin{equation}
	S(z_1)S(z_2) = e^{\frac{1}{4}\ln\left(\frac{1 + t_1 t_2^*}{1 + t_1^* t_2}\right) (\hat{b}^\dag \hat{b} + \hat{b}\hat{b}^\dag)} S(z_3)\, , 
\end{equation}
where we recall that $z_j = r_j \, e^{i \theta_j }$.

%
\subsection{Link to the $J$ coefficients} \label{app:link:composition:Js}
%

To derive  ~\eqref{eq:compact:squeezing} and~\eqref{eq:squeezing:new:coeffs} we first note that, for the combination of $\rm{exp}\left[  - i \, J_+ \, \hat B_+^{(2)} \right]$ and $\rm{exp}\left[ - i \, J_- \, \hat B_-^{(2)} \right]$, we have
\begin{align}
&r_1 = 2 \, J_+, && \theta_1 = \pi/2 \, , \nonumber \\
 &r_2 =  2 J_- , && \theta_2 =  \pi  \, .
\end{align}
These values can now be used to derive the coefficients $\varphi_J$ and $\zeta_J$. From~\eqref{app:eq:identities:for:tj} it follows that
\begin{equation}
\tanh(r_3) \, e^{i \theta_3}  = \frac{i\tanh(2J_+) - \tanh(2J_-)}{1-i\tanh(2J_+)\tanh(2J_-)} \, .
\end{equation}
The phase factor, defined as $ e^{ - i \, \varphi_J\, \hat N_b } $ above for the rotation can be derived in a similar manner. We first note that 
\begin{align}
\ln \left( \frac{1 + t_1 t_2^*}{1 + t_1^* t_2} \right) =  i \rm{Arg}\left( \frac{1 + t_1 t_2^*}{1 + t_1^* t_2} \right) = i \rm{Arg}\left( \frac{1 - i \tanh(2J_+) \tanh(2J_-) }{1 + i  \tanh(2J_+) \tanh(2J_-) }\right)  \, , 
\end{align}
which follows from the definition of the complex logarithm  and from the fact that $(1 + t_1 t_2^*)/(1 + t_1^* t_2)$ has complex norm $1$. The expression can now be simplified to 
\begin{align}
i \ \rm{Arg} \left(\frac{[1 - i \tanh(2 J_+) \tanh(2 J_-)]^2}{ 1 + \tanh^2(J_+) \tanh^2(J_-) } \right) = & \, 2 \, i\,  \rm{Arg} \left( 1 - i \tanh(2 J_+) \tanh(2J_-) \right)  \, ,
\end{align}
where the last equality follows from the fact that the angle in complex space matters, not the magnitude of the real and imaginary parts. Furthermore, we have that $\rm{Arg}(z^n) = n \, \rm{Arg}(z)$, which means that a factor of 2 can be pulled down in front of the expression.  Finally, we note that the Arg function is related to the atan2 function, a standard operation in many numerical libraries by the relation $\rm{Arg}(x + i y) = \rm{atan2}(y,x)$. However, if $x >0$, we find the special case $\rm{Arg}( x + i y) = \rm{arctan}(y/x)$. In our case, $x = 1$, and thus we find 
\begin{equation}
\varphi_J = \arctan \left( \tanh(2 J_+) \tanh(2 J_-) \right) \, ,
\end{equation}
where we have also accounted for a minus sign in the phase. 
These expressions can now be used to interpret the evolution induced by the mechanical single-mode squeezing term as a combination of a rotation and a squeezing, as discussed in the main text.

%
\subsection{Link between the $J$ coefficients to the Bogoliubov coefficients}\label{app:rel}

To obtain the relation between the functions $J_b$, $J_+$, and $J_-$ and the $P_{11}$ and $I_{P_{22}}$ functions, we remember that
\begin{equation}
	S_{sq} \hat{\mathbb{X}} = \left(\begin{array}{cc} \alpha & \beta \\ \beta^* & \alpha^* \end{array}\right)\hat{\mathbb{X}} \, ,
\end{equation}
and attempt to make it equivalent to
\begin{eqnarray}\label{eq:doublesqueeze}
\boldsymbol{S}_{R}(J_b) \boldsymbol{S}_\mathrm{sq}(2J_+,\pi/2) \boldsymbol{S}_\mathrm{sq}(2J_-,\pi)  & = \begin{pmatrix} S_{11} & S_{12} \\ S_{21} & S_{22} \end{pmatrix} \, , 
\end{eqnarray}
where we find  analogously to our result in~\eqref{app:eq:squeezing:elements:first} that the matrix elements are given by 
\begin{align} \label{app:eq:squeezing:elements}
\alpha = S_{11} &=e^{-iJ_b}\left(\cosh(2J_+)\cosh(2J_-) - i \sinh(2J_+)\sinh(2J_-)\right) \, , \nonumber \\
\beta = S_{12} &=  -e^{-iJ_b}\left( i \sinh(2J_+)\cosh(2J_-) -  \cosh(2J_+)\sinh(2J_-) \right)  \, .
\end{align}
A particular set of solutions to these equations is given as
\begin{align}\label{app:eq:squeezing:relation}
	J_+ = & \frac{\mathrm{arcosh}(|\alpha^2 - \beta^2|)}{4} \, ,\nonumber \\
	J_- = & \frac{1}{4} \mathrm{arcosh}\left(\frac{(2|\alpha|^2 -1)}{|\alpha^2 - \beta^2|}\right) \, ,\nonumber \\
	J_b = &  -\frac{1}{2} \mathrm{Arg}\left(\frac{\alpha^2 - \beta^2}{|\alpha^2 - \beta^2|}\right) \, .
\end{align}
We arrived at the expression for $J_b$ since 
\begin{equation}
e^{- 2 i J_b } = \left(\frac{\alpha^2 - \beta^2}{|\alpha^2 - \beta^2|}\right) \, .
\end{equation}
Taking the logarithm of a complex number gives  $\ln z = \ln |z| + i
\rm{Arg} z$, with $\rm{Arg}$ defined as in
Sec.~\ref{app:link:composition:Js}, and where $z \in
\mathbb{C}$. In this case, $|e^{ - 2 i J_b}|=1$, 
which means that we arrive at the expression above.

It is then straight-forward to relate the $J$ coefficients to $P_{11}$ and $I_{P_{22}}$ by using the expressions in~\eqref{app:eq:bogoliubov:expressions}.

\section{Derivation of the Fisher information}\label{appendix:sensing}

In this appendix we derive the QFI for estimation of an arbitrary parameter $\theta$ contained in the nonlinear Hamiltonian~\eqref{main:time:independent:Hamiltonian:to:decouple:dimensionless}. 
According to~\eqref{definition:of:QFI}, the QFI is obtained as
\begin{align}\label{definition:of:QFI:appendix}
\mathcal{I}_\theta
=& 4\sum_n \lambda_n\left(\bra{\lambda_n}\mathcal{\hat H}_\theta^2\ket{\lambda_n} - \bra{\lambda_n}\mathcal{\hat H}_\theta\ket{\lambda_n}^2 \right) - 8\sum_{n\neq m}\frac{\lambda_n \lambda_m}{\lambda_n+\lambda_m}\left| \bra{\lambda_n}\mathcal{\hat H}_\theta \ket{\lambda_m}\right|^2\;,
\end{align}
where the operator $\mathcal{\hat H}_\theta$ is defined as $\mathcal{\hat H}_\theta = - i \hat U^\dag_\theta \partial_\theta \hat U_\theta $. In order to emphasize that the time-evolution operator depends on the parameter $\theta$, we added the subscript. 

\subsection{Derivation of the coefficients}

Now we derive the expression for the QFI~\eqref{eq:main:result:QFI}. The commutators which appear in the calculation are listed in~\eqref{app:eq:commutators}. 
The operator $\mathcal{\hat H}_\theta$ has the form 
\begin{align} \label{app:eq:mathcalH:with:coefficients}
 \mathcal{\hat H}_\theta=A\,\hat N_a^2 +B\,\hat N_a +  C_+\,\hat B_+ + C_{\hat N_a,+} \hat N_a \, \hat B_+ + C_-\,\hat B_- + C_{\hat N_a,-} \, \hat N_a \, \hat B_- + E\,\hat N_b +  F\,\hat B_+^{(2)} + G\,\hat B_-^{(2)} +K. 
\end{align}
This is a consequence of the fact that the Lie algebra of the whole Hamiltonian is closed and finite. 

Let us proceed to determine the coefficients in~\eqref{app:eq:mathcalH:with:coefficients}. To do so, we first  differentiate the time-evolution operator $\hat U_\theta$ with respect to the parameter $\theta$. The operator $\hat U_\theta$ can be decomposed into the form
$\hat U_\theta=\hat U_{\hat N_a} \hat{\tilde U}_{\rm{sq}}\hat U_{\hat B_+} \hat U_{\hat B_-}$, where we have introduced
\begin{align}
 \hat U_{\hat N_a}
&= e^{-i\left( \Omega_{\rm{c}}  \, \tau +\mathcal{\hat F}_{\hat N_a}\right){\hat N}_a}\;,\A\\
\hat U_{\hat B_+}
&= e^{-i \, \mathcal{\hat F}_+ \, \hat B_+}\;,\A\\
\hat U_{\hat B_-}
&= e^{ - i  \, \mathcal{\hat F}_- \, \hat B_-}  \, ,
\end{align}
and where we recall that $\mathcal{\hat F}_{\hat N_a} = F_{\hat N_a}+F_{\hat N_a^2}\hat N_a$, 
$\mathcal{\hat F}_+ = F_{\hat B_+} + \, F_{\hat N_a \, \hat B_+} \hat N_a$, 
and $\mathcal{\hat F}_- = F_{\hat B_-} + \, F_{\hat N_a \,  \hat
  B_-}\hat{N}_a$.
To simplify notation, the differential operator $\partial_\theta$ is
understood in this section to act on the first symbol on its right
only. 
Then, we can write $\mathcal{\hat H}_\theta$ as
\begin{align}\label{H}
\mathcal{\hat H}_\theta
=& -i\left( 
\hat U_{{ \hat N_a}}^\dagger \partial_\theta {\hat U}_{{ \hat N_a}} 
+ \hat U_{\hat B_-}^\dagger \hat U_{\hat B_+}^\dagger {\hat {\tilde{U}}}_{\rm{sq}}^\dagger \partial_\theta{ \hat {\tilde{U}}}_{\rm{sq}} \hat U_{\hat B_+}\hat U_{\hat B_-} +\hat U_{\hat B_-}^\dagger \hat U_{\hat B_+}^\dagger \partial_\theta{\hat U}_{\hat B_+} \hat U_{\hat B_-}+ \hat U_{\hat B_-}^\dagger  \partial_\theta \hat{U}_{\hat B_-}
\right).
\end{align}
In order to proceed we need to compute the derivative
$\partial_\theta\hat{\tilde U}_{\rm{sq}}$, which requires us to
decompose the operator $\hat{\tilde U}_{\rm{sq}}$ as
in~\eqref{app:ansatz}, which we reprint here as $\hat{\tilde
  U}_{\rm{sq}}=  \exp[-i J_b \hat{N}_b]\,\exp[-i J_+\hat
B_+^{(2)}]\,\exp[-iJ_-\hat B_-^{(2)}]$, where $J_b$ and $J_\pm$ are
time-dependent real functions. We present the exact form of these
coefficients in~\eqref{eq:diff:equations:Js} in
Appendix~\ref{appendix:mechanical} as a solution to a coupled set of
differential equations.  If we now assume all three coefficients $J_b,
J_+$, and $J_-$ to depend on the estimation parameter $\theta$, we differentiate $\hat{\tilde{U}}_{\rm{sq}}$  to find 
\begin{align} \label{eq:app:dff:of:Us}
\partial_\theta \hat {\tilde{U}}_{\rm{sq}} =& \,   -   i \,  \partial_\theta J_b  \, \hat N_b \, e^{-i \,  J_b   \, \hat N_b}\,e^{-i  \, J_+ \, \hat B^{(2)}_+ }\,e^{- i \, J_- \, \hat B^{(2)}_-} - i  \, \partial_\theta J_+ \,  e^{-i \, J_b \, \hat N_b}\,   \hat B^{(2)}_+ \,  e^{-i \, J_+\, \hat B^{(2)}_+}\,e^{- i \, J_- \, \hat B^{(2)}_-} \A \\
& - i \, \partial_\theta J_-  \, e^{-i  \, J_b \, \hat N_b}\,e^{- i \,  J_+ \, \hat B^{(2)}_+}\, \hat B^{(2)} _- \, e^{ - i \, J_-\hat B^{(2)}_-} \;.
\end{align}
We then obtain $\hat U^\dagger_{\hat B_-}\hat U^\dagger_{\hat B_+} \hat{\tilde{U}}_{\rm{sq}}^\dagger\partial_\theta\hat{\tilde{U}}_{\rm{sq}}\hat U_{\hat B_+}\hat U_{\hat B_-} = \hat C_1+\hat C_2+\hat C_3$ with
\begin{align}
\hat C_1=
& - i \partial_\theta J_b \, \hat U_{\hat B_-}^\dag \hat U_{\hat B_+}^\dag  \, e^{i \, J_- \, \hat B^{(2)}_-}\,e^{i  \, J_+ \, \hat B^{(2)}_+ } \,  \hat N_b \, e^{-i  \, J_+ \, \hat B^{(2)}_+ }\,e^{- i \, J_- \, \hat B^{(2)}_-} \hat U_{\hat B_+} \hat U_{\hat B_-} \A \nonumber \\
 =&  - i \partial_\theta J_b \biggl[ \cosh(4 J_+) \cosh(4 J_-) \biggl(  \hat N_b + \hat B_+\,\hat{ \mathcal{F}}_- +\hat{\mathcal{F}}_-^2\, -\hat B_-\,\hat{\mathcal{F}}_++\hat{\mathcal{F}}_+^2\,  \biggr) \nonumber \\
& \quad\quad\quad\quad+ \frac{1}{2} \cosh(4 J_+) \sinh(4J_-) \biggl( \hat B_+^{(2)} +2\,\hat B_+ \,\hat{\mathcal{F}}_-+2\,\hat{\mathcal{F}}_-^2  \, +2\,\hat B_-\,\hat{\mathcal{F}}_+-2\,\hat{\mathcal{F}}_+^2\,\biggr) \nonumber \\
&\quad\quad\quad\quad +  \cosh(4 J_+) \sinh^2(2 J_-) + \sinh^2(2 J_+)- \frac{1}{2} \sinh(4 J_+) \biggl( \hat B_-^{(2)} + 2\,\hat B_-\, \hat{\mathcal{F}}_- -2\,\hat B_+\hat{\mathcal{F}}_{+}  - 4\,\hat{\mathcal{F}}_-\,\hat{\mathcal{F}}_+ \biggr)  \biggr] \, \nonumber ,
\\
\hat C_2=
 &- i  \partial_\theta J_+ \hat U_{\hat B_-}^\dagger \hat U_{\hat
   B_+}^\dagger e^{iJ_-\hat B_-^{(2)}}\hat B_+^{(2)}e^{-iJ_-\hat B_-^{(2)}}\hat U_{\hat B_+} \hat U_{\hat B_-}  \A \\
 =& - i \partial_\theta J_+  \biggl[ \cosh(4J_- ) \biggl(   \hat B_+^{(2)} +2\,\hat B_+ \,\hat{\mathcal{F}}_-+2\,\hat{\mathcal{F}}_-^2 \,  +2\,\hat B_-\,\hat{\mathcal{F}}_+ -2\,\hat{\mathcal{F}}_+^2\, \biggr) \nonumber \\
&\quad\quad\quad\quad + 2 \sinh(4 J_-) \biggl(  \hat N_b + \hat B_+\, \hat{\mathcal{F}}_-  + \hat{\mathcal{F}}_-^2\, - \hat B_-\,\hat{\mathcal{F}}_++\hat{\mathcal{F}}_+^2\,  \biggr) + \sinh(4 J_-) \biggr] \,, \nonumber \\
\hat C_3=
&- i\,  \partial_\theta J_- \hat U_{\hat B_-}^\dag \hat U_{\hat B_+}^\dag \, \hat B_-^{(2)} \hat U_{\hat B_+} \hat U_{\hat B_-} \nonumber \\
=& - i \, \partial_\theta J_-  \, \biggl[  \hat B_-^{(2)}  + 2\, \hat B_-\, \hat{\mathcal{F}}_- -2\, \hat B_+\hat{\mathcal{F}}_+  - 4\,\hat{\mathcal{F}}_-\,\hat{\mathcal{F}}_+  \biggr]\, .
\end{align}
For the remaining terms in $\hat{\mathcal{H}}$, we obtain
\begin{align}
 \hat U_{ \hat N_a}^\dagger \partial_\theta {\hat U}_{ \hat N_a}  &= - i\left( \tau \partial_\theta \Omega_{\rm{c}} +  \partial_\theta  \hat{\mathcal{F}}_{N_a}\right)  \hat N_a \;,  \nonumber \\
 \hat U_{\hat B_-}^\dagger  \partial_\theta \hat{ U}_{\hat B_-} &=  - i \partial_\theta \hat{\mathcal{F}}_{-} \hat B_- \;, \nonumber \\
\hat U_{\hat B_-}^\dagger \hat U_{\hat B_+}^\dagger \partial_\theta{\hat U}_{\hat B_+} \hat U_{\hat B_-} 
&=  - i \partial_\theta \hat{\mathcal{F}}_{+}  \left( \hat B_+ + 2\,\hat{\mathcal{F}}_-\right) \;.
\end{align}
By comparing the obtained expression for $\mathcal{\hat H}_\theta$ with the form~\eqref{app:eq:mathcalH:with:coefficients}, we find for the coefficients 
\begin{align}\label{qfi:coefficients}
A=& 
-\partial_\theta F_{\hat N_a^2}-2 F_{\hat N_a \, \hat B_-}\partial_\theta F_{\hat N_a\, \hat B_+}  + 2 F_{\hat N_a\, \hat B_-}F_{\hat N_a \, \hat B_+}R_{\partial_\theta, 0} + \sum_{s\in\{+,-\} } s \,e^{-s4J_-} F_{\hat N_a \, \hat B_s}^2 \, R_{\partial_\theta, s} 
\;,  \nonumber\\
B=&  -\tau\partial_\theta\Omega_{\rm{c}} - \partial_\theta F_{\hat N_a}-2 \, F_{\hat B_-}\partial_\theta F_{\hat N_a \, \hat B_+}-2 \, F_{\hat N_a\, \hat B_-}\partial_\theta F_{\hat B_+}
+ 2 \left(F_{\hat B_+}F_{\hat N_a\, \hat B_-}+F_{\hat B_-}F_{\hat N_a \, \hat B_+}\right)R_{\partial_\theta, 0}  \nonumber\\
&
+\sum_{s\in\{+,-\} } 2s e^{-s 4J_-}F_{\hat B_s}F_{\hat N_a \, \hat B_s} \, R_{\partial_\theta, s}
\; ,\nonumber\\
 C_\pm = & \, -\partial_\theta F_{\hat B_\pm } \pm  \,  F_{\hat B_\pm } R_{\partial_\theta, 0}
- e^{\pm 4J_-}  \, F_{\hat B_\mp} \, R_{\partial_\theta, \mp}
\;,\nonumber\\
C_{\hat N_a,\pm}=& \, - \partial_\theta  \, F_{\hat N_a \, \hat B_\pm} \pm  \,  F_{\hat N_a \, \hat B_\pm} \, R_{\partial_\theta, 0}
-e^{\pm 4J_-} \, F_{\hat N_a \, \hat B_\mp } \, R_{\partial_\theta, \mp}
\;,\nonumber\\
E = & \,  -\left(e^{4J_-} R_{\partial_\theta,-} - e^{-4J_-} R_{\partial_\theta,+} \right)/2  \;,\nonumber \\	
F = & \, -\left( e^{4J_-} R_{\partial_\theta,-} + e^{-4J_-} R_{\partial_\theta,+}\right)/4 \;,\nonumber \\
G = & \, - R_{\partial_\theta,0}/2   \;,\nonumber \\
K = & \, - 2   F_{\hat B_-} \,  \partial_\theta F_{\hat B_+} + 2 F_{\hat B_-} F_{\hat B_+} R_{\partial_\theta, 0}
+  \sum_{s\in\{+,-\}} s\, e^{-s4J_-} F_{\hat B_s}^2  \, R_{\partial_\theta, s} \,   + \, \partial_\theta J_b /2  + E/2 \, ,
\end{align}
where
\begin{align}
R_{\partial_\theta, 0} &= 2 \, \partial_\theta J_- - \sinh(4 J_+) \, \partial_\theta J_b \, , \nonumber \\
R_{\partial_\theta, \pm} &=  2 \,  \partial_\theta J_+ \mp  \cosh(4 J_+) \, \partial_\theta J_b \, .
\end{align}
The coefficients $E$ and $K$ will cancel out in the expression for $\mathcal{I}_\theta$, but we include them here for completeness. 

It is clear from the expressions above that the expressions simplify dramatically when the parameter $\theta$ to estimate is not contained in the coefficients $J_\pm$ and $J_b$, such that $\partial_\theta J_b = \partial_\theta J_\pm = 0$. For that case, we have $E=F=G=0$. 

\subsection{Derivation of the QFI expression}
The next step in the derivation of~\eqref{eq:main:result:QFI} is to take the expectation values of $\mathcal{\hat H}_\theta$ according to~\eqref{definition:of:QFI}. 
In order to do so, we will need the expectation values listed in Appendix~\ref{appendix:Fisher:nonlinear:commutators}. 
Noticing that the coefficients $E$ and $K$ will not contribute to the QFI, we drop them. Then we obtain
\begin{align}
 \braket{\lambda_n|\mathcal{\hat H_\theta}^2|\lambda_n} -\braket{\lambda_n|\mathcal{\hat H_\theta}|\lambda_n}^2
=&  A^2\left(4|\mu_{\rm{c}}|^6+6|\mu_{\rm{c}}|^4+|\mu_{\rm{c}}|^2\right)
+2AB\left(2|\mu_{\rm{c}}|^4+|\mu_{\rm{c}}|^2\right)
+B^2|\mu_{\rm{c}}|^2\nonumber \\
  &+(2n+1)\sum_{s\in\{+,-\}}\left(
    C_s^2+2C_sC_{\hat N_a,s}|\mu_c|^2+C_{\hat N_a,s}^2\left(|\mu_{\rm{c}}|^4+|\mu_{\rm{c}}|^2\right)\right)\nonumber\\
& +2(F^2+G^2)\left(n^2+n+1\right)\;,
\end{align}
and
\begin{align}
\left. \left| \braket{\lambda_n|\mathcal{\hat H}_\theta|\lambda_m} \right|^2  \right|_{n\neq m}
=&  \left(
\left(C_++C_{\hat N_a,+}
|\mu_{\rm{c}}|^2
\right)^2 
+ \left(C_-+C_{\hat N_a,-}|\mu_{\rm{c}}|^2\right)^2
\right)\left((m+1)\delta_{n,m+1}+m\delta_{n,m-1}\right)\A\\
&+ \left(F^2 +G^2\right)
\left((m+1)(m+2)\delta_{n,m+2}+m(m-1)\delta_{n,m-2}\right)\;, 
\end{align}
which can be written as 
\begin{align}
\left. \left| \braket{\lambda_n|\mathcal{\hat H}_\theta|\lambda_m} \right|^2  \right|_{n\neq m}
=& \left(
\left(C_++C_{\hat N_a,+}|\mu_{\rm{c}}|^2
\right)^2 
+ \left(C_-+C_{\hat N_a,-}|\mu_{\rm{c}}|^2\right)^2
\right)\left((m+1)\delta_{n,m+1}+(n+1)\delta_{m,n+1}\right)\A\\
&+ \left(F^2 +G^2\right)
\left((m+1)(m+2)\delta_{n,m+2}+(n+1)(n+2)\delta_{m,n+2}\right)\;.
\end{align}
where we changed the 
  summation index in the last term. We obtain that 
\begin{align}
\sum_{n\neq m}\frac{\lambda_n \lambda_m}{\lambda_n+\lambda_m}\left| \bra{\lambda_n}\mathcal{\hat H}_\theta \ket{\lambda_m}\right|^2 = 2\sum_{n}\left(\frac{\lambda_n \lambda_{n+1}}{\lambda_n+\lambda_{n+1}} C_{1\mathcal{H}}   + \frac{\lambda_n \lambda_{n+2}}{\lambda_n+\lambda_{n+2}}  C_{2\mathcal{H}} \right) \, ,
\end{align}
where
\begin{align}
C_{1\mathcal{H}} =& \left(
\left(C_-+C_{N_a,-}|\mu_{\rm{c}}|^2
\right)^2 
+ \left(C_-+C_{N_a,-}|\mu_{\rm{c}}|^2\right)^2\right)(n+1) \, ,\A \\
C_{2\mathcal{H}} =& \left(F^2  +  G^2\right)(n+1)(n+2)\;.
\end{align}
Using that $\lambda_n = \frac{\tanh^{2n}(r_T)}{\cosh^2(r_T)}$ and evaluating the sum in~\eqref{definition:of:QFI:appendix}, we obtain the result~\eqref{eq:main:result:QFI}.

\section{Coefficients and quantum Fisher information expressions}\label{app:coeff}
Our paper is based on general techniques for decoupling the Hamiltonian~\cite{Wei_1963,Bruschi:Xuereb:2018}. These techniques can be applied for any functional time-dependent behavior of the  parameters of the Hamiltonians; however, explicit result can be obtained only in the case that a specific form of the time dependence is specified.

In the main text, we argued that we are interested in the following forms of the couplings: $\tilde{\mathcal{G}}(\tau)=\tilde{g}_0(1+\epsilon\sin(\Omega_g\,\tau))$,  $\tilde{\mathcal{D}}_1(\tau)=\tilde{d}_1\,\cos(\Omega_{d_1}\,\tau)$,  and $\tilde{\mathcal{D}}_2 (\tau) = \tilde{d}_2 \, \cos(\Omega_{d_2})$. Here we will compute the $F$ functions~\eqref{sub:algebra:decoupling:solution} for the coupling expressions we have chosen.  Whenever $\tilde{\mathcal{D}}_2(\tau) = 0$, we find that $\xi=\exp[-i\,\tau]$.

\subsection{Coefficients for a time-dependent nonlinear coupling}\label{app:coeff:time:dependent:c1}
Here we list the coefficients for the dynamics when $\tilde{\mathcal{G}}(\tau) = \tilde{g}_0 ( 1 + \epsilon \,  \sin( \Omega_g \tau)  )$ and $\tilde{\mathcal{D}}_1(\tau) = \tilde{\mathcal{D}}_2(\tau) = 0$.  
\begin{align} \label{eq:time:dependent:coefficients}
F_{\hat{N}_a^2} &= -\tilde{g}_0^2
\bigl[ \tau -\sin (\tau) \cos (\tau) \bigr] +2 \, \epsilon \frac{\tilde{g}_0^2}{ \Omega_g }  \biggl[ \sin ^2( \tau ) \cos (\Omega_g \tau   )- 2 \sin^2 \left(\frac{\tau}{2} \right)\biggr]- \epsilon \frac{\tilde{g}_0^2}{ \Omega_g ( 1 + \Omega_g)} \,   \sin (2\tau ) \sin ( \Omega_g \, \tau )\nonumber \\
&- \epsilon \frac{4\tilde{g}_0^2}{ \Omega_g ( 1 - \Omega_g^2)} \,  \cos (\tau ) \sin^2\left( \frac{(1 - \Omega_g)\tau}{2} \right)
+ \epsilon^2 \, \frac{\tilde{g}_0^2}{ 4 \, \Omega_g ( 1 + \Omega_g)} \, \left( 2 \, \tau-4 \sin (\tau) \cos ( \Omega_g \, \tau ) (\cos (\tau) \cos ( \Omega_g \, \tau )-2)\right)  \nonumber \\
&+ \epsilon^2 \, \frac{\tilde{g}_0^2}{4 \, \Omega_g ( 1 - \Omega_g^2)} \, \biggl( 4 \, \sin (\tau) \cos ( \Omega_g \, \tau ) (\cos (\tau) \, \cos ( \Omega_g \, \tau )-2)+8 \cos (\tau) \,  \sin ( \Omega_g \, \tau )\nonumber \\
&\quad\quad\quad\quad\quad\quad\quad\quad\quad\quad+(1-2 \, \cos (2\, \tau)) \sin (2\,  \Omega_g \, \tau ) - 2 \, \tau\biggr) \nonumber \\
&+ \epsilon^2 \, \frac{\tilde{g}_0^2}{2 \, \Omega_g \, (1 - \Omega_g^2)^2} \biggl( 4 \, \Omega_g \,   \sin (\tau)\,  \cos ( \Omega_g \, \tau )- \Omega_g\,  \sin (2 \, \tau) \, \cos (2 \, \Omega_g \, \tau ) -4 \cos (\tau) \sin (\Omega_g \, \tau)+\cos (2 \, \tau) \, \sin (2 \, \Omega_g \, \tau ) \biggr) \,,\nonumber \\
F_{\hat N_a \, \hat{B}_+} &= -\frac{\tilde{g}_0}{1 + \Omega_g} \, \epsilon \sin(\tau) \sin(\Omega_g \, \tau) + \frac{2 \, \Omega_g \, \tilde{g}_0}{1 - \Omega_g^2} \, \epsilon \, \sin^2 \left( \frac{( 1 - \Omega_g)\tau}{2} \right) -\tilde{g}_0 \, \sin (\tau)\,,\nonumber \\
F_{\hat N_a \, \hat{B}_- } &=  - \frac{\tilde{g}_0 }{1 - \Omega_g} \, \epsilon \,  \sin (\tau) \, \cos ( \Omega_g \, \tau) + \frac{\tilde{g}_0}{1 - \Omega_g^2} \, \epsilon \, \sin((1 + \Omega_g)\tau)  - 2 \, \tilde{g}_0 \, \sin^2 \left( \frac{\tau}{2} \right) \, .
\end{align}
At resonance with $\Omega_g = 1$, these coefficients are given by 
\begin{align}\label{eq:resonance:coefficients}
F_{\hat N_a^2} &=-\frac{1}{16}  \tilde{g}_0^2 \, \bigl[ 16 \, \tau-8 \sin (2\, \tau)+\epsilon \, (32-36 \cos (\tau)+4 \cos (3 \, \tau))+\epsilon ^2 \,   \bigl( 6 \, \tau-4 \sin (2\, \tau)+\sin (2 \, \tau)\,\cos (2 \, \tau) \bigr)\bigr]\,,\nonumber\\
F_{\hat N_a \, \hat B_+} &= -\tilde{g}_0 \sin (\tau) \left(1+ \frac{\epsilon}{2} \sin (\tau)\right) \,,\nonumber \\
F_{\hat N_a \, \hat B_-} &=\frac{\tilde{g}_0}{4} \epsilon   \,  \left(  \sin (2 \, \tau)-2 \, \tau  \right) -  2  \, \tilde{g}_0 \,  \sin^2\left( \frac{\tau}{2} \right) \, .
\end{align}
Given these coefficients, the QFI for a general frequency $\Omega_g$ is given by 
\begin{align} \label{app:QFI:eq:g0:general:Omega}
\mathcal{I}_{\tilde{g}_0} = 
& \frac{4 \, \tilde{g}_0^2}{\Omega_g^2 \, ( \Omega_g^2 - 1)^4} \left| \mu_{\rm{c}}\right| ^2 \left(4 \, \left| \mu_{\rm{c}}\right| ^4+6 \,  \left| \mu_{\rm{c}}\right| ^2+1\right) \nonumber \\
&\quad\times \biggl(2 \,  \tau \, \Omega_g^5-4 \, \tau \, \Omega_g^3+2 \,  \tau \, \Omega_g- \tau \, \Omega_g^3 \epsilon ^2+\frac{1}{2} \Omega_g^2 \, \epsilon ^2 \sin (2 \,  \Omega_g \, \tau)+2 \,  \Omega_g^2 \,  \epsilon ^2 \,  \cos (\tau) \,  \sin ( \Omega_g \, \tau) \nonumber \\
&\quad\quad\quad+\tau \, \Omega_g  \, \epsilon ^2-4 \, \Omega_g^4 \,  \epsilon \, \cos (\tau) \sin ^2(\Omega_g \, \tau/2)-2 \, \left(\Omega_g^2-1\right) \Omega_g \sin (\tau) \left(\Omega_g^2-\epsilon \,  \sin ( \Omega_g \, \tau)-1\right) \nonumber \\
&\quad\quad\quad+4 \, \Omega_g^2 \, \epsilon   \, \cos (\tau) \, \sin ^2(\Omega_g \, \tau/2)-\epsilon \, \cos ( \Omega_g \, \tau) \left(2 \, \Omega_g^3 \, \epsilon \,  \sin (\tau)+\epsilon   \, \sin ( \Omega_g \, \tau)+2\,  \Omega_g^4-6 \,  \Omega_g^2+4\right) \nonumber \\
&\quad\quad\quad+2\, \Omega_g^4 \, \epsilon -6 \,  \Omega_g^2\, \epsilon +4\,  \epsilon \biggr)^2 \, \nonumber \\
&+  \, 4 \,  |\mu_{\rm{c}}|^2 \, \cosh(2 \,  r_T) \left( 1 + \frac{|\mu_{\rm{c}}|^2 }{\cosh^2 ( 2 \,  r_T)} \right) \nonumber \\
&\quad\times \biggl[ \left( 1 - \cos(\tau) - \epsilon \frac{\Omega_g \cos(\Omega_g \tau) \sin (\tau) - \cos(\tau) \sin(\Omega_g \, \tau) }{\Omega_g^2 - 1} \right)^2 \nonumber \\
&\quad\quad\quad+ \left( \sin( \tau) + \epsilon \frac{\Omega_g (1 - \cos(\tau)  \cos(\Omega_g \, \tau)) - \sin (\tau) \sin(\Omega_g \,  \tau) }{\Omega_g^2 - 1} \right)^2 \biggr]  \,  .
\end{align}
At resonance, the QFI becomes 
\begin{align} \label{eq:app:QFI:c1:res}
\mathcal{I}_{\tilde{g}_0}^{(\rm{res})} =& \frac{1}{16} \left| \mu_{\rm{c}}\right| ^2 \biggl[\tilde{g}_0^2 \left(4 \left|  \mu_{\rm{c}}\right| ^4+6 \left|  \mu_{\rm{c}}\right| ^2+1\right) \nonumber \\
&\quad\quad\quad\times \left(4 \tau \epsilon ^2-3 \epsilon ^2 \sin (2 \, \tau )-8 \, \tau \,  \epsilon  \sin (\tau)-32 \,  \epsilon  \cos (\tau)+ \, 2 \epsilon  (\tau \, \epsilon +2) \cos (2 \, \tau)+16 \, \tau -16 \sin (\tau)+28 \epsilon \right)^2 \nonumber \\
&+16 \cosh (2 r_T) \left(\left| \mu_{\rm{c}}\right| ^2 \frac{1}{\cosh^2(2 r_T)} +1\right) \left(\sin ^2(\tau) (\epsilon  \sin (\tau)+2)^2+(\tau \epsilon -\cos (\tau) (\epsilon  \sin (\tau)+2)+2)^2\right)\biggr] \, .
\end{align}

\subsection{Coefficients for a time-dependent linear displacement }\label{app:coeffs:time:dependent:d1}
We here print the $F$ coefficients for a time-dependent linear displacement term $\tilde{\mathcal{D}}_1(\tau) = \tilde{d}_1 \, \cos(\Omega_{d_1} \, \tau)$ and a constant light--matter coupling $\tilde{\mathcal{G}}(\tau) \equiv \tilde{g}_0$: 
\begin{align} \label{eq:app:F:coefficients:d1}
F_{\hat N_a} &=-\tilde{g}_0 \, \tilde{d}_1 \, \frac{\ 2 \Omega_{d_1}^2 \cos ^2(\tau) \,  \sin (\Omega_{d_1} \, \tau)+\sin (\Omega_{d_1} \, \tau) \left(\Omega_{d_1}^2 \, \cos (2\,  \tau )-3 \, \Omega_{d_1}^2+4\right)-4 \Omega_{d_1} \sin (\tau) \cos (\tau) \cos (\Omega_{d_1} \tau )}{2 \Omega_{d_1} \left(\Omega_{d_1}^2-1\right)}\,, \nonumber \\
F_{\hat N_a^2} &=\frac{1}{2} \tilde{g}_0^2 \left( \sin (2 \, \tau)-2 \, \tau \right)\,,\nonumber \\
F_{\hat B_+} &=  - \tilde{d}_1 \, \frac{\Omega_{d_1} \,  \cos (\tau)  \, \sin (\Omega_{d_1} \, \tau )-\sin (\tau)  \, \cos (\Omega_{d_1} \, \tau )}{1 - \Omega_{d_1}^2}\,,\nonumber \\
F_{\hat B_-} &=  - \tilde{d}_1 \, \frac{\Omega_{d_1} \,  \sin (\tau ) \sin (\Omega_{d_1} \, \tau)+\cos (\tau) \,  \cos (\Omega_{d_1} \, \tau )-1}{1 - \Omega_{d_1}^2}\,, \nonumber \\
F_{\hat N_a \, \hat B_+} &= -\tilde{g}_0 \, \sin (\tau)\,,\nonumber \\
F_{\hat N_a \, \hat B_-} &=\tilde{g}_0 \,  (\cos (\tau)-1)  \, .\nonumber \\
\end{align}
This yields the following expression for the QFI:
\begin{align}\label{result:formula:oscillating:d1}
\mathcal{I}_{\tilde{d}_1} 
=& \,  \frac{4 }{\Omega_{d_1}^2\left(1-\Omega_{d_1}^2\right)^2} \biggl[4 \, \tilde{g}_0^2  \, |\mu_{\rm{c}}|^2 \,  \left(\sin (\Omega_{d_1} \, \tau ) \,  \left(\Omega_{d_1}^2 (1-\cos (\tau )) - 1\right)+\Omega_{d_1} \,  \sin (\tau ) \,  \cos (\Omega_{d_1} \, \tau)\right)^2 \nonumber \\
&+ \frac{\Omega_{d_1}^2}{\cosh(2  \, r_T)} \bigl(2 +\left(\Omega_{d_1}^2-1\right) \sin^2 (\Omega_{d_1} \, \tau)-2\,  \Omega_{d_1} \sin (\tau) \sin ( \Omega_{d_1} \, \tau)-2 \, \cos (\tau) \cos ( \Omega_{d_1} \, \tau)\bigr)\biggr] \, .
\end{align}
For the constant case $\Omega_{d_1}=0$, we find
\begin{align}
\mathcal{I}_{\tilde{d}_1}^{\rm{(con)}} =& 16 \,\biggl(\tilde{g}_0^2 \left| \mu_{\rm{c}} \right| ^2 (\tau-\sin (\tau))^2 +\frac{\sin ^2\left(\tau/2\right)}{\cosh(2 \, r_T)} \biggr) \, .
\end{align}
At resonance with $\Omega_{d_1} = 1$, the coefficients become
\begin{align}
F_{\hat N_a} &=-\frac{1}{4} \tilde{g}_0 \, \tilde{d}_1 \left( \sin (3 \, \tau)-7 \sin (\tau)+4  \, \tau  \cos (\tau) \right)  \, ,\nonumber \\
F_{\hat N_a^2} &= - \frac{1}{2} \tilde g_0^2 (2\tau - \sin(2\tau)) \, ,  \nonumber\\
F_{\hat B_+} &= \frac{1}{2} \tilde{d}_1 \,  (\tau +\sin (\tau ) \,  \cos (\tau)) \nonumber \\
F_{\hat B_-} &= \frac{1}{2} \tilde{d}_1 \,  \sin^2(\tau) \, ,\nonumber \\
F_{\hat N_a \hat B_+} &= -\tilde g_0 \sin(\tau) \, , \nonumber \\
F_{\hat N_a \hat B_-} &= \tilde g_0 (\cos(\tau) - 1) \, ,\, 
\end{align}
and the Fisher information becomes
\begin{align} \label{eq:app:QFI:d1:res}
\mathcal{I}_{\tilde{d}_1}^{(\rm{res})} =&   4 \, \tilde{g}_0^2 \, |\mu_{\rm{c}}|^2 \left(\tau+\sin(\tau)\left(\cos(\tau)-2\right)\right)^2   +\frac{1}{\cosh{(2 r_T)}} \left(\tau^2+2\tau\sin(\tau)\cos(\tau)+\sin^2(\tau)\right) \, .
\end{align}

\subsection{Approximate coefficients for a constant and resonant squeezing}\label{app:coeffs:time:dependent:d1}
 In this section, we consider constant and time-dependent squeezing. The perturbative solutions to the time-dependent squeezing dynamics are only valid for $\tilde{d}_2 \ll 1$. For consistency, we will assume $\tilde{d}_2 \ll 1$ throughout this appendix, even for estimation of a constant squeezing strength. This assumption will also significantly simplify the expressions that follow. 

\subsubsection{Constant squeezing} \label{app:sec:constant:squeezing}

When we consider constant squeezing, i.e. $\Omega_{d_2}=0$ with $\tilde{\mathcal{D}}_2(\tau) \equiv \tilde{d}_2$, 
we find $\xi = \cos( \sqrt{1 + 4 \tilde{d}_2} \tau) + \sin(\sqrt{1 + 4 \tilde{d}_2}\tau)/\sqrt{1 + 4 \tilde{d}_2}$. 
For $\tilde{d}_2 \ll 1$ and $\tilde{d}_2 \tau \sim 1$, this expression
approximates to $\xi = e^{-i(1+2\tilde{d}_2)\tau}$. With the addition
of a constant light--matter coupling $\tilde{\mathcal{G}}(\tau) \equiv
\tilde{g}_0 $, the nonvanishing $F$ coefficients are 
(with $\tilde{D}_1=0$) 
\begin{align} 
F_{\hat N_a^2} &= -\tilde{g}_0^2 \,\frac{ 2(1 + 2\tilde{d}_2)\tau - \sin(2(1 + 2\tilde{d}_2)\tau) }{2(1+2\tilde{d}_2)^2}  \, ,\nonumber\\
F_{\hat N_a \hat B_+} &= -\tilde{g}_0 \, \frac{\sin((1 + 2\tilde{d}_2)\tau) }{1+2\tilde{d}_2} \, ,\nonumber \\
F_{\hat N_a \hat B_-} &= -\tilde{g}_0 \, \frac{1 - \cos((1 + 2\tilde{d}_2)\tau) }{1+2\tilde{d}_2} \, .
\end{align}
To simplify the expressions further we assume $0\ll\tilde{d}_2\ll\tilde{d}_2\tau\ll 1$ and discard terms proportional to $\tilde{d}_2$, while keeping only terms proportional to $\tilde{d}_2\tau$. We obtain
\begin{align} \label{app:eq:F:coeffs:constant:d2}
F_{\hat N_a^2} &= -\tilde{g}_0^2 \,\frac{ 2(1 + 2\tilde{d}_2)\tau - \sin(2(1 + 2\tilde{d}_2)\tau) }{2} \, , \nonumber\\
F_{\hat N_a \hat B_+} &= -\tilde{g}_0 \,\sin((1 + 2\tilde{d}_2)\tau) \, ,  \nonumber \\
F_{\hat N_a \hat B_-} &= -\tilde{g}_0 \, (1 - \cos((1 + 2\tilde{d}_2)\tau)) \, .
\end{align}
With the same approximations, and by using the relations~\eqref{app:eq:squeezing:relation}, we obtain
\begin{align} \label{app:eq:J:coeffs:constant:d2}
J_+ &=  0 \,,\quad J_- =  0 \quad\rm{and}\quad J_b =  (1 + 2\tilde{d}_2)\tau \, .
\end{align}
For this special case, many of the terms in the QFI
coefficients~\eqref{app:eq:squeezing:relation} are zero, $A = B=C_\pm
C_{\hat N_a, - } = G = F = 0$. 
The only nonzero coefficient is $C_{\hat N_a, +} = 2 \tilde{g}_0\tau$.
We then find the QFI
\begin{align} \label{eq:app:QFI:d2:con:app}
\mathcal{I}_{\tilde{d}_2}^{(\rm{const,app})} =   8 \,  \tilde{g}_0^2 \,  \tau^2  \, |\mu_{\rm{c}}|^2 \frac{1}{\cosh (2  \, r_T)} \left(1+ 2 |\mu_{\rm{c}}|^2+\cosh (4 \, r_T)\right) \, .
\end{align}

\subsubsection{Resonant time-dependent squeezing} \label{app:sec:resonant:squeezing:coefficients}

In the next step, we will consider the resonant case. Using the approximate solution for $\Omega_{d_2}=2$, which gives the expression of $\xi(\tau)$~\eqref{app:eq:RWA} and small $\tilde d_2$ given in~\eqref{app:eq:RWA} and neglecting all terms proportional to $\tilde d_2$ but keeping expressions proportional to $\tilde d_2 \tau$, we obtain for the nonvanishing  $F$ coefficients
\begin{align} \label{app:eq:F:coeffs:resonant:d2}
F_{\hat N_a^2} &= \tilde{g}_0^2 \, \frac{   \cosh(2 \tilde{d}_2 \tau)\sin(2\tau) + \sinh(2 \tilde{d}_2 \tau) -2\tau }{2}  \, ,\nonumber\\
F_{\hat N_a \hat B_+} &= -\tilde{g}_0 \left( \cosh(\tilde{d}_2 \tau)\sin(\tau) + \sinh(\tilde{d}_2 \tau)\cos(\tau) \right) \, , \nonumber \\
F_{\hat N_a \hat B_-} &= \tilde{g}_0 \left( \cosh(\tilde{d}_2 \tau)\cos(\tau) + \sinh(\tilde{d}_2 \tau)\sin(\tau)  -1 \right) \, .
\end{align}
Furthermore, using the relations between $\alpha$ and $\beta$ and the $J$ coefficients in~\eqref{app:eq:squeezing:relation}, we find under the same approximations as above,
\begin{align} \label{app:eq:J:coeffs:resonant:d2}
J_+ &=  \frac{1}{2} \tilde{d}_2\tau \,,\quad J_- =  0 \quad\rm{and}\quad
J_b =  \tau \, .
\end{align}
We obtain for the QFI
\begin{align} \label{eq:app:QFI:d2:res:app}
\mathcal{I}_{\tilde{d}_2}^{(\rm{res,app})} =&   \,4 \tau^2\left( \tilde{g}_0^4 (4|\mu_{\rm{c}}|^6 + 6|\mu_{\rm{c}}|^4 + |\mu_{\rm{c}}|^2) + \tilde{g}_0^2 |\mu_{\rm{c}}|^2 \, \frac{|\mu_{\rm{c}}|^2 + \cosh(2r_T)^2}{\cosh(2r_T)} + \frac{\cosh^2(2r_T)}{\cosh^2(2r_T)+1} \right) \, .
\end{align}

\end{document}